# Kondo induced $\pi$-phase shift of microwave photons in a circuit quantum electrodynamics architecture

Guang-Wei Deng[1,2,3]†, Loïc Henriet[4,5]†, Da Wei[1,2], Shu-Xiao Li[1,2], Hai-Ou Li[1,2], Gang Cao[1,2], Guang-Can Guo[1,2], Marco Schiró[6], Karyn Le Hur[4], Guo-Ping Guo[1,2,*]

[1] Key Laboratory of Quantum Information, University of Science and Technology of China, Chinese Academy of Sciences, Hefei 230026, China

[2] Synergetic Innovation Center of Quantum Information & Quantum Physics, University of Science and Technology of China, Hefei, Anhui 230026, China

[3] Institute of Fundamental and Frontier Sciences, University of Electronic Science and Technology of China, Chengdu 610054, China

[4] CPHT, CNRS, Ecole Polytechnique, Institut Polytechnique de Paris, Route de Saclay, 91128 Palaiseau, France

[5] ICFO-Institut de Ciencies Fotoniques, The Barcelona Institute of Science and Technology, 08860 Castelldefels (Barcelona), Spain

[6] Institut de Physique Théorique, Université Paris Saclay, CNRS, CEA, F-91191 Gif-sur-Yvette, France

*Corresponding author: gpguo@ustc.edu.cn

†These authors contributed equally to this work.

**Abstract**

Mesoscopic systems constitute appealing platforms to study many-body physics with light and matter degrees of freedom. The Kondo effect refers to the screening of a spin-1/2 impurity by a cloud of conduction electrons, then forming a many-body Fermi liquid ground state. The Kondo resonance produces a phase shift in the transmitted electronic wave packet which depends on the symmetry and nature of the many-body ground state. Theoretical calculations suggest that the Kondo resonance can interact with the irradiation photon field and should give rise to a π-phase shift of the photon signal in the case where the ground state is a Fermi liquid. This π-phase shift of microwave photon is driven from the Korringa-Shiba relation of quantum impurity Fermi-liquid ground states. We report the first observation of such a π-phase shift in a graphene double quantum dot within a circuit quantum electrodynamics architecture where the microwave photons couple to the pseudo-spin or charge degrees of freedom. The observed Kondo temperature $T_K \sim 550$ mK is in agreement with DC conductance measurements. All our results support the formation of a Kondo resonance located above the Fermi level of the electronic reservoirs and the occurrence of an SU(4) Fermi-liquid ground state. We finally study how the Kondo-photon interactions can be tuned by inter-dot electron tunnel coupling strengths. Our experimental achievements may contribute to a better understanding of many-body physics in hybrid circuit systems, and open up new applications in atomic thin materials from the light-matter interaction.

**Introduction**

A lot of efforts have been carried out to study Kondo physics in quantum dot and mesoscopic geometries (*1-8*). The occurrence of a Kondo resonance at the mesoscale produces a phase shift in the transmitted electronic wave packet which depends on the symmetry and nature of the many-body ground state (*9-12*). Recently advances in circuit quantum electrodynamics (cQED) architectures have been used to detect charge or orbital degrees of freedom and reveal the effect of charge fluctuations in Kondo effect (*13-20*), which offer an appealing platform to explore many-body physics and light-matter interactions. Previous theory predicted that Kondo effect could be suppressed by external irradiation (*21*). Photon induced Kondo satellites in a single electron transistor (*22*), microwave induced charge frozen at Kondo resonance (*16*) and scaling laws of the Kondo problem at finite frequency (*20*) have been reported. However, the backaction effect of Kondo correlation on the irradiation field, such as the coupled microwave field, has not been explored yet.

Here we predict that a Kondo effect would give rise to a π-phase shift of a microwave photon signal at resonance in the case where the ground state is a Fermi liquid. This $\pi$-phase shift of microwave photon is driven from the Korringa-Shiba relation of quantum impurity Fermi-liquid ground states. In this case we consider a double quantum dot system (DQD), where the two dots are strongly capacitively coupled, which should then reveal SU(4) Kondo correlations with entangled spin and charge degrees of freedom (*23-26*). In the Kondo model, the spin or here the pseudo-spin related to the charge or orbital degrees of freedom on the two dots, is described by a dynamical susceptibility $\chi(\omega)$ which describes the response to an applied AC field. If one fixes the frequency of the AC field to be close to the resonance frequency of the many-body system $\omega_K$ such that $\hbar\omega_K = k_B T_K$, with $\hbar$ and $k_B$ being the Planck and the Boltzmann constants, then one obtains the important equality $J(\omega_K)\chi(\omega_K) = i$ (*27*) showing the Fermi liquid ground state formation (*25, 28*). Here, $J(\omega)$ represents the spectral function associated to the transmission channel which transports the microwave signal. This relation which holds for $\omega = \omega_K$ implies a form of dynamical spin susceptibility which agrees with the Korringa-Shiba relation of quantum impurity Fermi liquid systems at low frequencies (*29, 30*). This relation is known to show and lead to important predictions for many-body Kondo physics and charge fluctuations in quantum RC circuits at low frequencies (*31-34*). Here, we generalize the analysis to the regime $\omega\sim\omega_K$ by coupling to the microwave channel or a transmission line. If we couple directly the pseudo-spin to the transmission line, the reflection coefficient of the microwave photon $S_{11}(\omega) = Ae^{i\phi} = 1 + 2iJ(\omega)\chi(\omega)$ shows a $\Delta\phi = \pi$ phase shift for the resonance condition $\omega = \omega_K$ (*27, 35-37*). This $\pi$ phase shift of microwave photons should occur when fixing the photon frequency $\omega$ in resonance with the many-body frequency $\omega_K$, assuming the ground state is a quantum-impurity Fermi liquid system.

It is relevant to emphasize that for an open transmission line, $\Delta\phi = 0$ (*38*). This shows that many-body physics is important in the present case and directly measurable from the photon phase shift. This phase of the reflected light signal is also different from the Friedel's phases $\delta = \pi/2$ and $\pi/4$ which have been observed in the DC electronic conductance $2e^2/\hbar\, sin^2\delta$ for the SU(2) (*1-3, 12*) and SU(4) Kondo effects (*4-7, 39*), respectively.

Below, we report the observation of such a $\pi$ phase shift of microwave photon by coupling a graphene DQD with a microwave resonator. Our study of microwave photon response supported by DC conductance measurements across the mesoscopic electron system agrees with an SU(4) Kondo model which shows entangled spin and charge degrees of freedom (*23, 24, 26*), and with an interacting Breit-Wigner resonance (associated to transport) located above the Fermi energy evolving with temperature and bias voltage. In addition, a Kondo effect that was measured in graphene with point defects (vacancies) (*40*), has not yet been reported in graphene quantum-dot systems (*41, 42*). To show the robustness of the results, we present two cooling procedures. We will also show the effect of tuning (increasing) the tunneling strength between the two dots. In Ref. (*14*), low-frequency measurements of microwave light in carbon nanotube systems were performed in accordance with theory, and the authors reported a phase in radians which is not too small, still measured away from resonance such that the $\pi$ phase value was not attained. To analyze the data, we build a microscopic model describing the coupling of the mesoscopic DQD system to the cQED architecture. This will then give rise to a quantized mode model or an harmonic oscillator with a bare frequency $\omega_0$ transporting the microwave power. The capacitive light-matter coupling will result in a many-body shift of this frequency $\omega_0$, leading to a resonant condition called $\omega_0^*$ and englobing the formation of the Kondo resonance related to $T_K$.

Regarding the description of the electronic reservoirs, we consider that the DQD size is much larger than the atomic distance between neighboring sites in graphene, such that only the symmetric wave function associated to sub-lattices A and B points of the honeycomb lattice (shown in Fig. 1D) couple to the mesoscopic system. This leads to a one-channel description in each electronic reservoir in addition to the spin flavor, as in the SU(4) Kondo effect (*23*). We assume that the chemical potential does not lie at the neutrality point, to have a finite density of states in the reservoirs such that the Kondo effect can occur (*43*).

Our experimental findings are well supported by theoretical calculations, and pave the way for Kondo physics and its applications in atomic thin materials.

## Results

**Sample introduction**. The device shown in Fig. 1 is mounted in a dilution refrigerator, its base temperature is about 30 mK. Our previous works on graphene double quantum dots were performed both in series (*44*) and in parallel (*45*) shapes, allowing us to tune with precision the double dot orbital levels with DC gates. Two DQDs, made of few-layer etched graphene, are coupled to the resonator through their sources (*46*), however, only one of them is used while the other is grounded throughout the course of present experiment. Transport measurements are performed through source and drain (Fig. 1B), and the charge stability diagram is achieved by adjusting the gate voltages $V_{LP}$ and $V_{RP}$ (Fig. 1D). The electronic temperature of the DQD at base temperature is estimated to be about 100 mK, which has been reported by a temperature dependence experiment (*47*). A vector network analyzer (VNA) is used to apply coherent microwave driving tone and measure the reflection signal $S_{11}$ (defined from the scattering matrix, see Fig. 1C). The reflection signal can be measured by its amplitude ($A$=$|S_{11}|$) and phase ($\phi$=arg($S_{11}$)) components through the VNA. The resonance frequency of the resonator is 6.35 GHz, with a quality factor of ~3000. Details of sample fabrication and measurement setup are shown in the Supplementary Information Sec. I).

**Measuring the Kondo effect in a graphene double quantum dot.** A dispersive readout measurement permits to obtain the charge stability diagram of the DQD system (see Fig. 1D) (*47*), which shows the equilibrium charge state of the two dots as a function of the gate voltages $V_{LP}$ and $V_{RP}$. This diagram exhibits a typical hexagonal structure (*48*). At point D, the electronic configuration (*N,M*) corresponds to *N* electrons in the left dot and *M* electrons in the right dot. Along the grey lines (points A, B and C), two charge configurations are degenerate. By maintaining the electrochemical potential of one dot, we establish the Coulomb diamond of the other dot in Fig. S1 in the Supplementary Information. From this method, we find the charging energy of both dots to be $E_C$~0.4 meV, and from the charge stability diagram the mutual capacitance between dots is comparable to the single dot capacitances (*46*). Below, we study in more detail the charge degeneracy point A, where the pseudo-spin ½ will correspond to the two degenerate charge states (*N*-1,*M*) and (*N*,*M*-1), with one state on each quantum dot.

We measure the differential conductance d$I$/d$V$ at point A and two peaks at $V_{\rm sd} \sim \pm 0.05$ mV are observed (Fig. 1E). These two peaks constitute a first signature of the emergence of an SU(4) Kondo resonance. Generalizing the Fermi-liquid approach (*28*) to the SU(4) Kondo effect (*23, 25, 26*), it is indeed predicted that the conductance across the DQD at small bias voltages essentially follows $(2e^2/\mathrm{h})k_B T_K \rho(\hbar\omega = e|V_{sd}|)(|t'|/k_B T_K)^2$ where $k_B$ is the Boltzmann constant, and $t' \sim 0.3\, k_B T_K$ describes the tunnel coupling strength between the quantum dots which can be obtained from the microwave response (*47*), $T_K$ is the Kondo energy scale and $\rho$ is the spectral function associated to the Kondo resonance within the Fermi-liquid description (*23, 26*). The presence of anomalies at $V_{sd} \sim \pm 0.05$ mV are consistent with a spectral function $\rho$ described by a resonance at $\epsilon_0 = eV_{\rm sd} \cong k_B T_K$ above the Fermi energy of the reservoir electron leads, as predicted by the theory (*9*). These two peaks in the d$I$/d$V$ electronic transport response occur for an electron energy $eV_{\rm sd} \sim 50$ μeV, which corresponds to a Kondo temperature $T_K \sim 550$ mK. The d$I$/d$V$ characteristics is useful to detect the SU(4) Kondo effect because it reflects the spectral function on the mesoscopic system which is known for the SU(4) Kondo effect to show peaks not at the Fermi level, but rather at the positions $\pm k_B T_K$ in bias voltages (*26*). The mesoscopic system then behaves similarly as an interacting resonant level system centered at the position $\epsilon_0 = k_B T_K$, with a width $\Gamma$ that will be studied below as a function of bias voltage and temperature and which could not be described in terms of a non-interacting resonant level model. We verify the Kondo effect by a temperature dependent experiment, shown in Fig. S3 in the Supplementary Information.

**Kondo induced robust $\pi$ phase shift of microwave photon.** Fixing gate voltages at point A, we measure the microwave photon response as a function of $V_{sd}$, in order to characterize the light-matter interactions. We underline that the microwave results are obtained at the renormalized resonance frequency $\omega_0^*(V_{sd})$, which is determined from the minimum of the amplitude $A=|S_{11}|$ as a function of frequency (see Fig. S2 in the Supplementary Information) (*42*). In Figs. 2(A,C,E), we present the results for the phase and amplitude of the microwave signal, showing a phase shift of the order of $\pi$ for a prominent window of bias voltages $V_{sd}$ applied across the mesoscopic electron system. In addition, the amplitude of the signal also reveals two dips at low energy scale, in agreement with that of Fig. 1E. As shown in Fig. 2A, a very robust phase of $\pi$ is observed from $V_{\mathrm{sd}}\cong 0$ to $V_{\mathrm{sd}}\cong \pm 0.4$ mV, above which the phase suddenly drops to zero. In Fig. 2C the amplitude (in dB) shows also two pronounced dips around $V_{\mathrm{sd}}\cong \pm 0.4$ mV. We notice that these features in the experimental results can be associated with an energy scale which is comparable with the charging energy $E_C$.

In the present geometry, the transmission line transporting the microwave power is described by the spectral function $J(\omega) = 2\alpha\hbar\omega\exp(-\omega/\omega_c)$ where the high-frequency cutoff $\omega_c$ is taken large compared to the driving frequency and the charging energy $E_C$. To fit the data, we assume that $\alpha{\sim}10^{-5}$; this value of $\alpha$ can also account for other internal dissipation effects (*49*). To evaluate the reflection coefficient $S_{11}$, we attach to the transmission line the cQED resonator (cavity), which is described by the photon field $\hat{x}$ characterizing the quantization of the electric field inside the cavity. From standard input-output theory (*50*), then we obtain $S_{11}(\omega, V_{sd}) = -1 + 2iJ(\omega)\chi_{\mathrm{xx}}^R(\omega, V_{sd})$, where the retarded dynamical susceptibility $\chi_{\mathrm{xx}}^R$ associated with the microwave field can be tuned as a function of the bias voltage $V_{\mathrm{sd}}$ applied across the mesoscopic electron system and frequency of the microwave power. We can write the general form:

$$\chi_{\mathrm{xx}}^{R} = \frac{\hbar\omega_0}{(\hbar\omega)^2-(\hbar\omega_0)^2-\hbar\omega_0\Pi^R(\omega)+iJ(\omega)\hbar\omega_0}, \qquad (1)$$

which includes the contributions coming from coupling the cQED resonator with the electronic system through $\Pi^{\mathrm{R}} = \mathrm{Re}(\Pi^{\mathrm{R}}) - i\mathrm{Im}(\Pi^{\mathrm{R}})$ and with the transmission line through $J(\omega)$. In particular, $\Pi^R$ will traduce the formation of the Kondo resonance between the DQD and the electronic reservoirs. Through Eq. (1) the resonance condition for light is defined as

$$(\hbar\omega_0^*(V_{sd}))^2 = (\hbar\omega_0)^2 + \hbar\omega_0 Re(\Pi^R(\omega_0^*(V_{sd})). \qquad (2)$$

Before describing in a quantitative manner the capacitive coupling between the resonator and the mesoscopic electron circuit, we can pursue at a general level the theoretical discussion on the occurrence of the π phase shift. More precisely, from Eqs. (1) and (2), we deduce that $\chi_{\mathrm{xx}}^{R}(\omega_0^*)\,[\mathrm{Im}\,\Pi^R(\omega_0^*) + \mathrm{J}(\omega_0^*)] = -i$, which is a generalization of the formula mentioned in the introduction for the spin susceptibility (the minus sign on the right-hand side in “$-i$” here traduces the minus sign in the definition of the photon susceptibility compared to the spin susceptibility given in introduction and the Kondo frequency is here replaced by the resonance light frequency). When $\Pi^R(\omega_0^*) = 0$, we check that $S_{11}$=1 in agreement with an open transmission line. On the contrary, if dissipation effects on the cQED resonator comes from the coupling with the mesoscopic electron system forming the Kondo resonance, then the condition Im $\Pi^R(\omega_0^*) > \mathrm{J}(\omega_0^*)$ will result in a π phase shift in $S_{11}$.

This phase shift is not yet a (complete) proof of the formation of a many-body Kondo resonance because at this stage this just traduces that the cQED system couples more strongly with another medium than the transmission line, here the mesoscopic electron system, characterized by a Breit-Wigner resonance with a Lorentzian form for the electron spectral function. The quantitative analysis developed below taking into account the renormalization effects in the width of this resonance, in agreement with the formation of a many-body Fermi liquid ground state, will justify the occurrence of such a $\pi$ phase shift of light until energies of the order of the charging energy. When increasing the bias voltage across the mesoscopic system, interaction-interaction effects in the Kondo resonance are well-known to produce decoherence effects which then increase considerably the width of the resonance with this bias voltage (*51*)**.** We will adapt this calculation of the resonance width and position to our system, showing that theory results of Figs. 2(B,D,F) are in agreement with an SU(4) Kondo effect. Neglecting many-body interaction effects, we could not explain these surprising results. Indeed, as soon as the bias voltage $V_{\mathrm{sd}}$ would become larger than the width of the non-interacting resonant level, which is usually much smaller than the charging energy to account for Coulomb blockade physics in the mesoscopic circuit, then dissipation effects from the mesoscopic circuit would again be very small generally implying that Im $\Pi^R(\omega_0^*)\sim 0$ then resulting in $S_{11}\sim 1$.

To build the Kondo model and our quantitative analysis for the light-matter coupling (see Sec. II of the Supplementary Information), we introduce in addition to the spin $S_z=\pm 1/2$ of an electron, the orbital pseudospin quantum number $T_z=\pm 1$ associated with the two degenerate charge states on the DQD. The 4 quantum states on the DQD coupled to the symmetric wavefunction made of the sub-lattices A and B in the graphene reservoir leads produce an SU(4) Kondo model (see Sec. III of the Supplementary Information). Within this description of the electronic sector, the light-matter coupling which describes the coupling between the resonator or cQED to the left lead in Fig. 1 can be written as a capacitive coupling to the orbital degree of freedom $\lambda T_z \hat{x}$, where $\lambda$ quantifies the coupling between the cavity or the resonator and the mesoscopic electron system. Studying the evolution of $\omega_0^*$ with respect to the driving frequency in Fig. S2, we deduce $\lambda/(\hbar\omega_0^*) = 10^{-2}$ and we adjust the value of $\omega_0^*$ for different voltages $V_{sd}$. To make a precise link with an interacting resonant level, it is useful to refermionize the pseudospin-1/2 particle: $T_z = 2(d^\dagger d - 1/2)$, $T^+ = d^\dagger$ and $T^- = d$, where the fermionic operators $d$ and $d^\dagger$ describe the orbital degrees of freedom on the DQD. The associated spectral function ρ to this d-particle is then shown in Eq. (11) of the Supplementary Information. A diagrammatic analysis then leads to Im $\Pi^R(\omega_0^*)\sim\lambda^2$ Im$\chi(\omega_0^*)$, where $\chi(\omega)$ is the spin susceptibility introduced earlier. The Korringa-Shiba relation for $\chi(\omega)$ then gives the estimate $\text{Im}\,\Pi^R\,(\omega_0^*)\sim \lambda^2\,\hbar\omega_0^*/T_K^2\sim 10^{-4}\,\hbar\omega_0^*$ which then justifies that at low bias voltages $\text{Im}\,\Pi^R\,(\omega_0^*) > J(\omega_0^*)$, resulting in the π-phase shift of light. More details on $\text{Im}\,\Pi^R$ can be found in Eq. (17) of the Supplementary Information.

Within the same diagrammatic approach, we then evaluate the bias-dependent resonant level width, and demonstrate in Sec. III. C of the Supplementary Information that the formulas in Ref. (*51*):

$$\Gamma = k_B T_K \text{ for } eV_{sd} \ll k_B T_K, \quad (3)$$

$$\Gamma \sim eV_{sd}/\ln^2(eV_{sd}/k_B T_K) \text{ for } eV_{sd} \gg k_B T_K \quad (4)$$

are valid for an SU(4) Kondo effect. At large bias voltages, the current enlarges the typical width of the Kondo resonance, which is the effect responsible for the robustness of the phase shift. In Sec. III. C of the Supplementary Information, we check that these forms are still valid for an SU(4) Kondo effect. Using this form of Γ, with a Kondo temperature $T_K \cong 550$ mK and a position $\epsilon_0 \sim k_B T_K$ of the Kondo resonance, we can explain (almost quantitatively) the experimental observations; see Figs. 2(B,D,F). The low bias features in the amplitude signal can also be accounted for by our model. Note that the low bias anomaly observed in the phase response in Fig. 2A is beyond the scope of our present Fermi-liquid analysis. Increasing the bias voltage such that $eV_{sd} > k_B T_K$, the logarithmic renormalization factors in Eq. (4) then contribute to maintain important dissipation effects until $\Gamma > e|V_{sd}|$. Using results of Fig. S3 of the Supplementary Information and Eqs. (3) and (4), then we predict that the π-phase shift in the reflected light signal subsists until $e|V_{sd}| \sim 10\ k_B T_K$ which means roughly the charging energy $E_c$. For larger bias voltages, when $e|V_{sd}| \gg \Gamma$ using Eq. (17) of the Supplementary Information and Fig. S3, we can then check that $\text{Im}\Pi^R\,(\omega_0^*)\sim 0$, justifying that the phase shift then smoothly drops to zero in the very high-bias regime.

A similar renormalization group argument can be formulated as a function of temperature showing that the π phase shift should indeed disappear when $T \sim T_K$. A detailed microscopic model of the experimental set up, the construction of the effective SU(4) Kondo model and its coupling to the photon field together with few relevant calculations are included in Sec. III of the Supplementary Information; where we also show that a bias-independent resonant level model (modeling the two degenerate charge states) would not explain the experimental data, showing that many-body effects through the formation of a Kondo resonance cannot be ignored. For completeness, in the Supplementary Information (see Fig. S5 and S6), we also show that the effect of the micro-wave field at point D can restore the same Kondo characteristics.

**Tuning the electron tunnel coupling strengths between quantum dots.** In Fig. 3, we show results for a different device, referred to as device 2, with different tunnel coupling strengths $t'$ for points A. Using similar methods as we discussed previously in device 1, we estimate the temperature scale $T_K$ of device 2 to be ~ 600 mK (corresponding to an energy scale of ~12.2 GHz). For points 1 and 2 in Fig. 3, essentially $t'<k_B T_K$, therefore we obtain similar results as in Fig. 2 and Fig. S3. In that case, $t'$ does not enter in the description of the light-matter coupling since the resonator couples directly to the pseudo-spin $T_z$ which is described by the formation of the Kondo energy scale $k_B T_K$. The phase shift observed in the microwave measurement in Fig. 3D progressively decreases from point 1 to point 3. For point 3, where $t'>k_B T_K$, the two dots are strongly coupled and form a molecule (large dot) which enters in a strongly Coulomb blockaded situation and almost decouple from the leads. The charge on the DQD becomes a fixed integer value and the conductance through the DQD becomes zero in the limit of $V_{sd}$~0. Equivalently, $t'$ acts as an orbital magnetic field along X direction for the pseudo-spin (see Sec. IIIA of the Supplementary Information), which gradually suppresses the Kondo effect such that the microwave does not reveal signatures of a π phase shift in that case.

## Conclusion

We have shown how the Kondo resonance can interact with the microwave photon field, producing a π-phase shift of light at low temperatures. The microwave frequency is adjusted carefully at the many-body light resonance frequency corresponding to the minimum of the amplitude of the light response, revealing important many-body dissipation effects. Applying the Korringa-Shiba relation of quantum impurity Fermi liquid Kondo models and renormalization group arguments to include bias voltage effects in the Breit-Wigner resonance and in the spectral function on the DQD, then we have shown how such a $\pi$-phase shift of light can subsist until energies larger than the Kondo scale, roughly around the charging energy of the mesoscopic system. Our findings are also in agreement with a many-body electronic spectral function which is located above the Fermi energy in agreement with an SU(4) symmetry in the Fermi-liquid description. Our work shows how microwave spectroscopy can probe the dynamical properties of the correlated electron system when the microwave frequency is synchronized with the typical frequency scale associated with the correlated phenomenon. Similar to three-lead or four-lead geometries (*8*), the microwave resonator allows then to study important features of the spectral function on the mesoscopic system. We have quantitatively described photon-electron many-body effects in our device and our work opens up new possibilities for fundamental and practical applications in many-body light-matter systems. An analogous microwave signature of charge Kondo effect could be observed in a hybrid metal-semiconductor implementation of a single-electron transistor (*52*).

**Author Contributions:** G.-W.D., D.W. designed and fabricated the samples. G.-W.D., S.-X.L., H.-O.L. and G.C. performed the measurements. L.H., M.S., and K.L.H. developed the theory analysis and calculations, and run the simulations. G.-W.D., L.H., M.S., K.L.H., and G.-P.G. analyzed the data and prepared the manuscript. G.-P.G., and G.-C.G. advised on experiments and data analysis. G.-W. D, K.L.H., L.H., M.S., G.-C.G and G.-P.G. wrote the manuscript with input from all the other authors.

**Acknowledgements:** We acknowledge discussions with Camille Aron, Sebastian Diehl, Serge Florens, Loïc Herviou, Takis Kontos, Vladimir Manucharyan, Christophe Mora, Nicolas Roch and Achim Rosch on related hybrid systems. We thank Olesia Dmytruk, Pascal Simon, Mircea Trif and Ryosuke Yoshii for useful comments. This work was supported by the National Key Research and Development Program of China (Grant No. 2016YFA0301700, and 2018YFA0306102), the National Natural Science Foundation of China (Grants Nos. 11625419, 61704164, 91836102, 61674132, 11674300, and 11575172), and the Anhui Initiative in Quantum Information Technologies (Grants No. AHY080000). This research has also benefitted from support from the DOE, under the grant DE-FG02-08ER46541. This work is partially supported by ANR BOCA. This work was partially carried out at the USTC Center for Micro and Nanoscale Research and Fabrication.

## References

1. D. Goldhaber-Gordon *et al.*, Kondo effect in a single-electron transistor. *Nature* **391**, 156-159 (1998).
2. W. G. van der Wiel *et al.*, The Kondo effect in the unitary limit. *Science* **289**, 2105-2108 (2000).
3. L. Kouwenhoven, L. Glazman, Revival of the Kondo effect. *Phys. World* **14**, 33-38 (2001).
4. P. Jarillo-Herrero *et al.*, Orbital Kondo effect in carbon nanotubes. *Nature* **434**, 484-488 (2005).
5. A. Makarovski, J. Liu, G. Finkelstein, Evolution of transport regimes in carbon nanotube quantum dots. *Phys. Rev. Lett.* **99**, 066801 (2007).
6. T. Hata *et al.*, Enhanced Shot Noise of Multiple Andreev Reflections in a Carbon Nanotube Quantum Dot in SU(2) and SU(4) Kondo regimes. *Phys. Rev. Lett.* **121**, 247703 (2018).
7. A. J. Keller *et al.*, Emergent SU(4) Kondo physics in a spin-charge-entangled double quantum dot. *Nature Phys.* **10**, 145-150 (2014).
8. R. Leturcq *et al.*, Probing the Kondo density of states in a three-terminal quantum ring. *Phys. Rev. Lett.* **95**, 126603 (2005).
9. Y. Ji, M. Heiblum, D. Sprinzak, D. Mahalu, H. Shtrikman, Phase Evolution in a Kondo-Correlated System. *Science* **290**, 779 (2000).
10. M. Zaffalon, A. Bid, M. Heiblum, D. Mahalu, V. Umansky, Transmission Phase of a Singly Occupied Quantum Dot in the Kondo Regime. *Phys. Rev. Lett.* **100**, 226601 (2008).
11. M. Sato, H. Aikawa, K. Kobayashi, S. Katsumoto, Y. Iye, Observation of the Fano-Kondo Antiresonance in a Quantum Wire with a Side-Coupled Quantum Dot. *Phys. Rev. Lett.* **95**, 066801 (2005).
12. S. Takada *et al.*, Transmission Phase in the Kondo Regime Revealed in a Two-Path Interferometer. *Phys. Rev. Lett.* **113**, 126601 (2014).
13. A. Dousse *et al.*, Controlled Light-Matter Coupling for a Single Quantum Dot Embedded in a Pillar Microcavity Using Far-Field Optical Lithography. *Phys. Rev. Lett.* **101**, 267404 (2008).
14. M. R. Delbecq *et al.*, Coupling a Quantum Dot, Fermionic Leads, and a Microwave Cavity on a Chip. *Phys. Rev. Lett.* **107**, 256804 (2011).
15. T. Frey *et al.*, Dipole Coupling of a Double Quantum Dot to a Microwave Resonator. *Phys. Rev. Lett.* **108**, 046807 (2012).
16. M. M. Desjardins *et al.*, Observation of the frozen charge of a Kondo resonance. *Nature* **545**,

71-74 (2017).
17. S. Léger *et al.*, Observation of quantum many-body effects due to zero point fluctuations in superconducting circuits. *Nature Commun.* **10**, 5259 (2019).
18. P. Forn-Díaz *et al.*, Ultrastrong coupling of a single artificial atom to an electromagnetic continuum in the nonperturbative regime. *Nature Phys.* **13**, 39 (2016).
19. L. Magazzù *et al.*, Probing the strongly driven spin-boson model in a superconducting quantum circuit. *Nature Commun.* **9**, 1403 (2018).
20. L. E. Bruhat *et al.*, Scaling laws of the Kondo problem at finite frequency. *Phys. Rev. B* **98**, 075121 (2018).
21. A. Kaminski, Y. V. Nazarov, L. I. Glazman, Suppression of the Kondo Effect in a Quantum Dot by External Irradiation. *Phys. Rev. Lett.* **83**, 384-387 (1999).
22. A. Kogan, S. Amasha, M. A. Kastner, Photon-Induced Kondo Satellites in a Single-Electron Transistor. *Science* **304**, 1293 (2004).
23. L. Borda, G. Zarand, W. Hofstetter, B. I. Halperin, J. von Delft, SU(4) Fermi liquid state and spin filtering in a double quantum dot system. *Phys. Rev. Lett.* **90**, 026602 (2003).
24. M.-R. Li, K. Le Hur, Double-Dot Charge Qubit and Transport via Dissipative Cotunneling. *Phys. Rev. Lett.* **93**, 176802 (2004).
25. C. Mora, P. Vitushinsky, X. Leyronas, A. A. Clerk, K. Le Hur, Theory of nonequilibrium transport in the SU(N) Kondo regime. *Phys. Rev. B* **80**, 155322 (2009).
26. K. Le Hur, P. Simon, D. Loss, Transport through a quantum dot with SU(4) Kondo entanglement. *Phys. Rev. B* **75**, 035332 (2007).
27. K. Le Hur, Kondo resonance of a microwave photon. *Phys. Rev. B* **85**, 140506 (2012).
28. P. Nozières, A “fermi-liquid” description of the Kondo problem at low temperatures. *J. Low Temp. Phys.* **17**, 31 (1974).
29. H. Shiba, The Korringa Relation for the Impurity Nuclear Spin-Lattice Relaxation in Dilute Kondo Alloys. *Prog. Theor. Phys.* **54**, 967-981 (1975).
30. M. Garst, P. Wölfle, L. Borda, J. von Delft, L. Glazman, Energy-resolved inelastic electron scattering off a magnetic impurity. *Phys. Rev. B* **72**, 205125 (2005).
31. C. Mora, K. Le Hur, Universal resistances of the quantum resistance–capacitance circuit. *Nature Phys.* **6**, 697-701 (2010).
32. M. Filippone, K. Le Hur, C. Mora, Giant Charge Relaxation Resistance in the Anderson Model. *Phys. Rev. Lett.* **107**, 176601 (2011).
33. J. Gabelli *et al.*, Violation of Kirchhoff's Laws for a Coherent RC Circuit. *Science* **313**, 499 (2006).
34. M. Büttiker, A. Prêtre, H. Thomas, Dynamic conductance and the scattering matrix of small conductors. *Phys. Rev. Lett.* **70**, 4114-4117 (1993).
35. M. Schiro, K. Le Hur, Tunable hybrid quantum electrodynamics from nonlinear electron transport. *Phys. Rev. B* **89**, 195127 (2014).
36. K. Le Hur *et al.*, Many-body quantum electrodynamics networks: Non-equilibrium condensed matter physics with light. *CR Phys.* **17**, 808-835 (2016).
37. K. Le Hur *et al.*, Driven dissipative dynamics and topology of quantum impurity systems. *CR Phys.* **19**, 451-483 (2018).
38. A. A. Clerk, M. H. Devoret, S. M. Girvin, F. Marquardt, R. J. Schoelkopf, Introduction to quantum noise, measurement, and amplification. *Rev. Mod. Phys.* **82**, 1155-1208 (2010).
39. T. Delattre *et al.*, Noisy Kondo impurities. *Nature Phys.* **5**, 208-212 (2009).

40. J. H. Chen, L. Li, W. G. Cullen, E. D. Williams, M. S. Fuhrer, Tunable Kondo effect in graphene with defects. *Nature Phys.* **7**, 535-538 (2011).
41. M. Eich *et al.*, Spin and Valley States in Gate-Defined Bilayer Graphene Quantum Dots. *Phys. Rev. X* **8**, 031023 (2018).
42. M. Raith, C. Ertler, P. Stano, M. Wimmer, J. Fabian, Electric control of tunneling energy in graphene double dots. *Phys. Rev. B* **89**, 085414 (2014).
43. J. Hopkinson, K. Le Hur, É. Dupont, Quantum dot in the pseudogap Kondo state. *Eur. Phys. J. B* **48**, 429-432 (2005).
44. L. J. Wang *et al.*, Gates controlled parallel-coupled double quantum dot on both single layer and bilayer graphene. *Appl. Phys. Lett.* **99**, 112117 (2011).
45. L. J. Wang *et al.*, Controllable tunnel coupling and molecular states in a graphene double quantum dot. *Appl. Phys. Lett.* **100**, 022106 (2012).
46. G. W. Deng *et al.*, Coupling two distant double quantum dots with a microwave resonator. *Nano Lett.* **15**, 6620- (2015).
47. G.-W. Deng *et al.*, Charge Number Dependence of the Dephasing Rates of a Graphene Double Quantum Dot in a Circuit QED Architecture. *Phys. Rev. Lett.* **115**, 126804 (2015).
48. W. G. van der Wiel *et al.*, Electron transport through double quantum dots. *Rev. Mod. Phys.* **75**, 1 (2002).
49. H. Max *et al.*, Spin-boson model with an engineered reservoir in circuit quantum electrodynamics. arXiv:1506.09114 (2015).
50. C. W. Gardiner, M. J. Collett, Input and output in damped quantum systems: Quantum stochastic differential equations and the master equation. *Phys. Rev. A* **31**, 3761 (1985).
51. A. Rosch, J. Kroha, P. Wolfle, Kondo effect in quantum dots at high voltage: Universality and scaling. *Phys. Rev. Lett.* **87**, 156802 (2001).
52. Z. Iftikhar *et al.*, Two-channel Kondo effect and renormalization flow with macroscopic quantum charge states. *Nature* **526**, 233-236 (2015).

# Figures

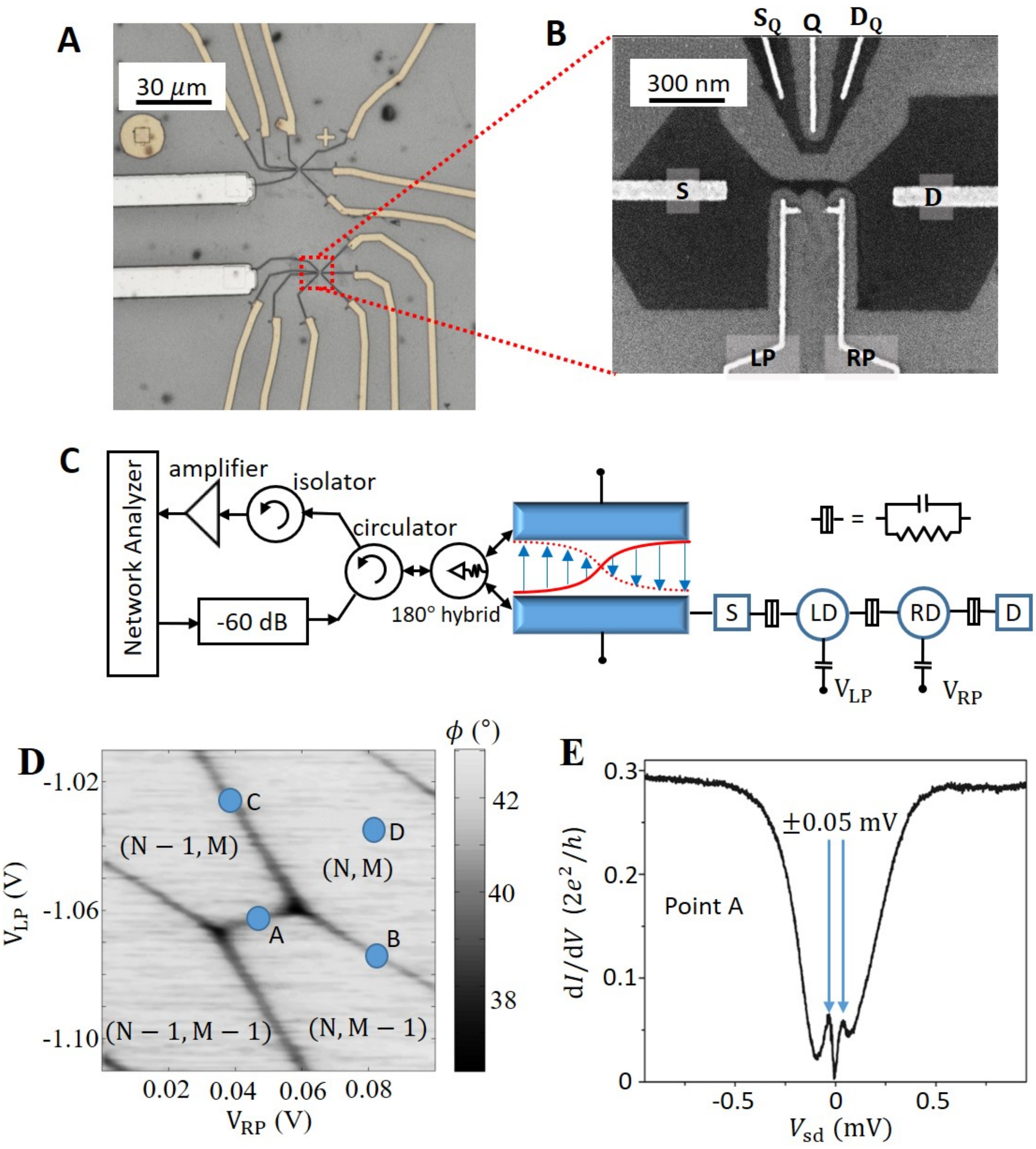

**Figure 1. Sample structure of the graphene DQD and transport peaks at small bias voltages.** (A) Micrograph of the DQD gate structure. (B) Sample structure of a typical etched graphene DQD. The dc voltages used to control the charge numbers in the DQD are applied via left and right plunger (LP and RP) gates. A quantum point contact with a source ($S_Q$) and drain ($D_Q$) channel and a tuning gate ($Q$) is integrated near the DQD. (C) Schematics of the hybrid device. The half-wavelength reflection line resonator is connected to DQD's left dot (LD) at one end of its two stripelines. The right dot (RD) is connected to the drain. A microwave signal is applied to the other end of the resonator, and the reflected signal is detected using a network analyzer. (D) Charge stability diagram of the DQD measured by dispersive readout with the resonator. (E) Differential

conductance at point A from panel D, showing two peaks at finite voltage $V_{\rm sd} \sim \pm 0.05$ mV.

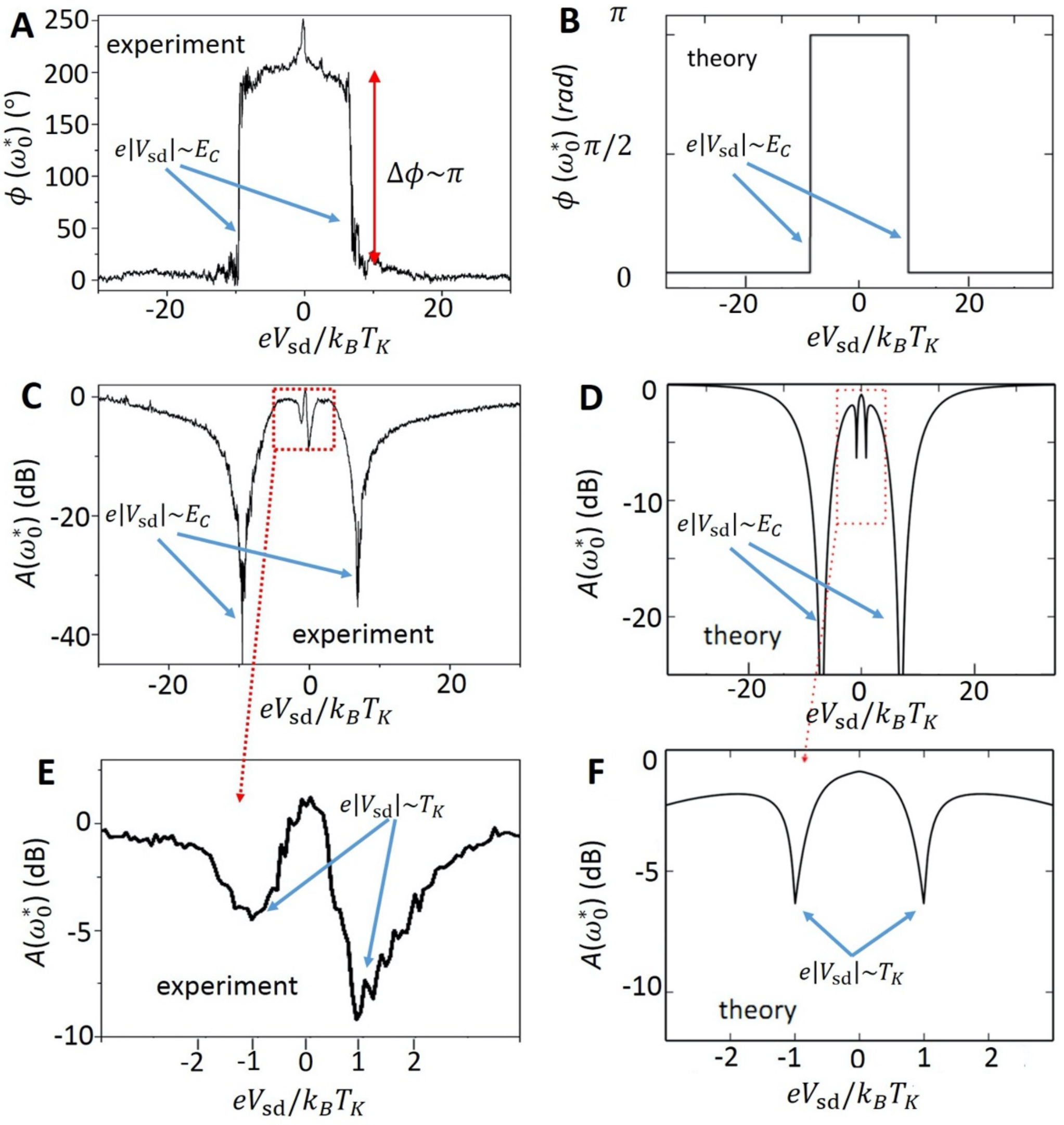

**Figure 2. Microwave measurements and theoretical results.** (A) Experimental result for the phase $\phi(\omega_0^*)$ of the reflected signal $S_{11}(\omega_0^*)$ at point A and at resonance $\omega = \omega_0^*(V_{sd})$ (determined experimentally, see fig. S2 in the Supplementary Information), and its evolution with respect to bias voltage. (B) Theory result for the phase based on the effective quantum impurity model. To fit the data, we extract the light-matter coupling from the frequency shift in Fig. S2, and find $\lambda/(\hbar\omega_0^*) = 10^{-2}$; we test different values of the dissipation parameter α and find that results are quite robust to different values of α; we choose $\alpha=10^{-5}$. The Kondo energy scale $T_K \cong 550$ mK is obtained from a fit of the low energy features. We find $t' \sim 0.3\, k_BT_K$ from the microwave response. The parameter Γ is derived from the Fermi-liquid theory together with Eqs. (3) and (4). (C) and (E) Experimental results for the amplitude $A$ of the reflection coefficient $S_{11}(\omega_0^*)$ at point A, as a function of the bias voltage; the zoom focuses on low-energy features. (D) and (F) Theory results based on the effective quantum impurity model.

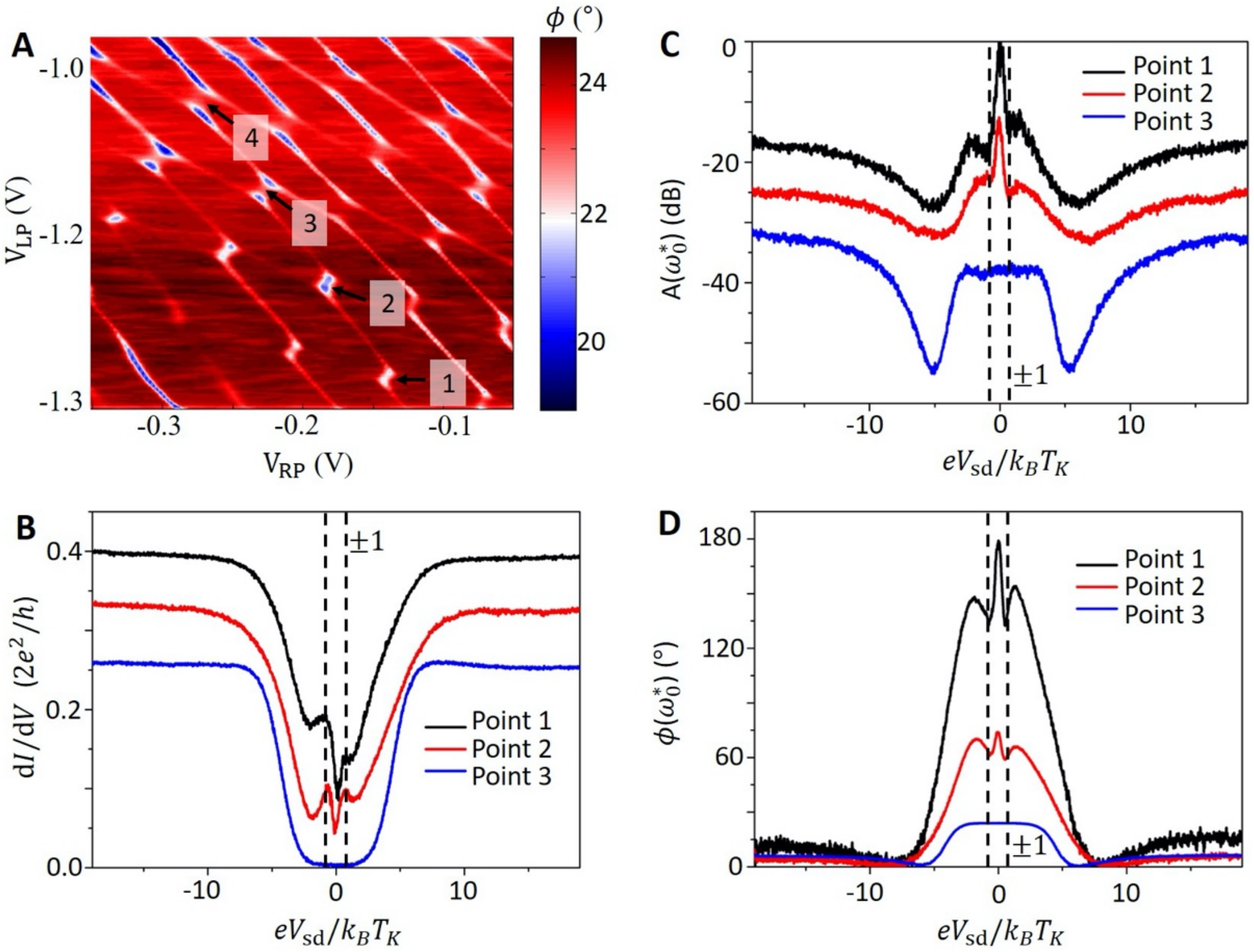

**Figure 3. Evolution of the results with inter-dot coupling strength $t'$ for a second device.** (A) Charge stability diagram for device 2. Inter-dot tunnel coupling strengths $t'$ of point 1, 2 and 3 are fitted to be 5.8 GHz, 8.4 GHz and 15.6 GHz, respectively, using a typical cavity response fitting procedure (*15, 47*). The point 4 shows a too large $t'$ parameter to be characterized by our cavity measurements. (B) Electronic differential conductance across the DQD for points 1,2,3. (C) Amplitude and (D) phase response of the microwave light signal as a function of the bias voltage $V_{sd}$ across the mesoscopic electron system for the three points 1, 2, 3. We observe that the phase shift of point 2 cannot reach 180 degree. Curves at points 1 and 2 in (B) have an offset of 0.5 and in (C) the offset is 10 dB.

# Supplementary Information for

## Kondo induced $\pi$-phase shift of microwave photons in a circuit quantum electrodynamics architecture

Guang-Wei Deng[1,2,3]†, Loïc Henriet[4,5]†, Da Wei[1,2], Shu-Xiao Li[1,2], Hai-Ou Li[1,2], Gang Cao[1,2], Guang-Can Guo[1,2], Marco Schiró[6], Karyn Le Hur[4], Guo-Ping Guo[1,2,*]

[1] Key Laboratory of Quantum Information, University of Science and Technology of China, Chinese Academy of Sciences, Hefei 230026, China

[2] Synergetic Innovation Center of Quantum Information & Quantum Physics, University of Science and Technology of China, Hefei, Anhui 230026, China

[3] Institute of Fundamental and Frontier Sciences, University of Electronic Science and Technology of China, Chengdu 610054, China

[4] CPHT, CNRS, Ecole Polytechnique, Institut Polytechnique de Paris, Route de Saclay, 91128 Palaiseau, France

[5] ICFO-Institut de Ciencies Fotoniques, The Barcelona Institute of Science and Technology, 08860 Castelldefels (Barcelona), Spain

[6] Institut de Physique Théorique, Université Paris Saclay, CNRS, CEA, F-91191 Gif-sur-Yvette, France

* Corresponding author. Email: gpguo@ustc.edu.cn

† These authors contributed equally to this work.

**The PDF file includes:**

**Supplementary Information Text**

**Sec. I Sample fabrication and measurement setup**

The samples are fabricated as follows. First, we mechanically exfoliated the graphene from its bulk, KISH graphite (Kyocera. Inc), to an undoped silicon chip with 285 nm oxide. In this experiment, we need two pieces of few-layer graphene with proper distance between 20 to 80 $\mu$m, and we selected those that met this requirement. Second, electron beam lithography (EBL) was employed several times, starting with the fabrication of alignment marks, then plasma-etching masks and electrode patterns. The EBL resists used were PMMA 950k A4 for the first step and double-layered PMMA 950k A2 for the latter two steps. We developed the sub-micrometer patterns under 0℃ to establish a better control of the device specifications. Through etching out all the undesired part of the graphenesheet to realize the designed device, we strove for the all-metal-side-gated configuration as described in Ref. [1], to avoid unstable gate terminals. This etching was carried out by inductively-coupled plasma (ICP), using a 4:1 gasmixture of Oxygen to Argon. For marks and electrodes, we deposited 5 nm Ti and 45 nm Au with an electron-beam evaporator. Finally, the resonator was fabricated by optical lithography followed by metal deposition in a thermalevaporator. The metal used was 200-nm-thick Al.

The microwave response was measured using a network analyzer (NA). The input and output ports of the NA were connected to the resonator via a circulator and a 180 degree hybrid, which splits the reflected signal back to the NA.Two 30 dB attenuators were connected between the NA output port and the circulator, reducing the power applied to the resonator down to lower than −130 dBm. The reflected signal was amplified first at 4 K and then at room temperature, producing an additional gain of 60 dB, and an isolator was used to prevent noise from the amplifiersand the environment from reaching the sample. The direct transport current was amplified by a low-noise currentpre-amplifier, before being measured by a digital multimeter.

The Coulomb diamond can be measured by transport experiment and revealing the charging energy $E_C$, as shown in Fig. S1.

**Sec. II Graphene DQD and light-matter coupling**

The Hamiltonian of a double dot setup takes the form [2, 3] $H_1 = H_L + H_R + H_D + H_T$, where

$$H_{l=L,R} = \sum_{\kappa\sigma} (\epsilon_\kappa + V_l) c^\dagger_{\kappa\sigma l} c_{\kappa\sigma l}$$

$$H_D = \sum_{l=L,R} \sum_{p_l\sigma} \left(\epsilon_{p_l} d^\dagger_{p_l\sigma} d_{p_l\sigma}\right) + \gamma_L n_L (n_L - 1) + \gamma_R n_R (n_R - 1) + \gamma_M n_L n_R$$

$$H_T = \sum_{l=L,R} \sum_{\kappa\sigma} \sum_{p_l} \left(t_{\kappa n_l} c^\dagger_{\kappa\sigma_l} d_{p_l\sigma} + h.c.\right) + \sum_{p_L p_R \sigma} \left(t_{p_L p_R} d^\dagger_{p_L\sigma} d_{p_R\sigma} + h.c.\right). \quad (1)$$

The first line describes the electrons in the Left (L) and Right (R) leads (respectively source and drain). $c_{\kappa\sigma l}$ ($c^\dagger_{\kappa\sigma l}$) is the annihilation (creation) operator of an electron of spin $\sigma$ and energy $\epsilon_\kappa$ and in the lead $l$, on which we apply a voltage $V_l$. $V_{sd} = V_L - V_R$ denotes the bias voltage applied across the device. The term $H_D$ describes the two dots: $d_{p_l\sigma}$ ($d^\dagger_{p_l\sigma}$) is the annihilation (creation) operator of an electron of spin $\sigma$ in the dot $l = L, R$ and at the energy level $p_l$. Here the coefficients $\gamma$ are functions of the capacitances of the tunnel junctions and $n_l = \sum_{p_l,\sigma} d^\dagger_{p_l\sigma} d_{p_l\sigma}$ denotes the number of electrons in the dot $l$. The part $H_T$ accounts for lead-dot and inter-dots tunneling events.

The model presented here can be explicitly refined to incorporate the precise geometry of the graphene lattice (leads) [4], and consider the two sub-lattices A and B of the honeycomb lattice. Since the DQD size (of the order of a few hundred nanometers) is much larger than the atomic distance between neighboring sites in graphene, we can assume that the tunneling probability from the leads in a given quantum dot is independent of the sub-lattice. In Eq. (1), $H_T$ can then be written as a function of the symmetric operator $1/\sqrt{2}(c_{\kappa\sigma lA} + c_{\kappa\sigma lB})$ in a given lead $l$. The kinetic term leading to a Dirac type linear dispersion can also be diagonalized in this basis if we assume a symmetric configuration for the A and B sites with respect to the direction of the tunneling. The use of either open or periodic boundary conditions in the orthogonal direction then lead to a one-dimensional kinetic Hamiltonian along the direction of the tunneling, which is diagonal in this symmetric/anti-symmetric basis [5, 6]. With these arguments, Hamiltonian (1) remains valid with the electronic operators referring to the symmetric operator (between sub-lattices A and B) in a given lead. Notice that $\kappa$ formally refers to the wave-vector along the direction of propagation across the mesoscopic system.

The interaction between microwave photons and the electronic degrees of freedom is often approximated by a dipolar coupling [8-11], the authors provided a general description in the case of a nano-circuit embedded inside the cavity. In our setup, the electrons of the left lead couple to the electric field at the extremity of a superconducting microwave resonator, as shown in Fig. 1(c). The total Hamiltonian of the coupled light-matter system becomes $H = H_1 + H_2$ with

$$H_2 = \hbar\omega_0 a^\dagger a + \frac{\lambda}{\sqrt{2}}(a + a^\dagger)(\textstyle\sum_{\kappa\sigma} c^\dagger_{\kappa\sigma L} c_{\kappa\sigma L}). \tag{2}$$

We apply one first unitary transformation on the Hamiltonian $H$ and define $H' = U^\dagger H U$. We have:

$$U = \exp\left[i\theta(\textstyle\sum_{\kappa\sigma} c^\dagger_{\kappa\sigma L} c_{\kappa\sigma L})\right], \tag{3}$$

where $\theta = \frac{\lambda(a-a^\dagger)}{i\sqrt{2}\hbar\omega_0}$. We can compute the transformed Hamiltonian $H'$ thanks to the Baker-Campbell-Haussdorff formula $e^X Y e^{-X} = Y + [X,Y] + \frac{1}{2!}\big[X,[X,Y]\big] + \frac{1}{3!}\Big[X,\big[X,[X,Y]\big]\Big] + \cdots$. We then get

$$U^\dagger c_{\kappa\sigma l} U = c_{\kappa\sigma l} e^{i\theta} \tag{4}$$

$$U^\dagger(\hbar\omega_0 a^\dagger a)U = -\frac{\lambda}{\sqrt{2}}(a+a^\dagger)(\textstyle\sum_{\kappa\sigma} c^\dagger_{\kappa\sigma L} c_{\kappa\sigma L}) \tag{5}$$

$$U^\dagger[\lambda(a+a^\dagger)(\textstyle\sum_{\kappa\sigma} c^\dagger_{\kappa\sigma L} c_{\kappa\sigma L})]U = \frac{\lambda^2}{\hbar\omega_0}(\textstyle\sum_{\kappa\sigma} c^\dagger_{\kappa\sigma L} c_{\kappa\sigma L})^2 \tag{6}$$

Assuming small fluctuations concerning the total number of conduction electrons, the right hand side of Eq. (6) is almost a constant, so that the resulting Hamiltonian reads $H' = H_L + H_R + H_D + H'_T + \hbar\omega_0 a^\dagger a$, where we have

$$H'_T = \textstyle\sum_{\kappa\sigma}\sum_{pL}(t_{\kappa pL} t^\dagger_{\kappa\sigma L} d_{pL\sigma} e^{-i\theta} + h.c.) + \sum_{\kappa\sigma}\sum_{pR}\big(t_{\kappa pR} t^\dagger_{\kappa\sigma R} d_{pR\sigma} e^{-i\theta} + h.c.\big) + \sum_{pLpR\sigma}(t_{pLpR} d^\dagger_{pL\sigma} d_{pR\sigma} + h.c.) \tag{7}$$

This unitary transformation has suppressed the explicit coupling between the source lead and the resonator. The tunneling terms from the source lead to the left dot have moreover acquired a phase which depends on the state of the resonator.

We then apply a second unitary transformation $\widetilde{H} = V^\dagger H' V$, with

$$V = \exp\,[i\theta(\textstyle\sum_{PL\sigma} d^\dagger_{pL\sigma} d_{pL\sigma})], \tag{8}$$

After similar calculations, the resulting Hamiltonian reads $\widetilde{H} = H_s + H_d + \widetilde{H}_D + \widetilde{H}_T + \hbar\omega_0 a^\dagger a$, where we have

$$\widetilde{H}_D = H_D - \frac{\lambda}{\sqrt{2}}(a+a^\dagger)(\sum_{pL\sigma} d^\dagger_{pL\sigma} d_{pL\sigma})$$

$$\widetilde{H}_T = \textstyle\sum_{l=L,R}\sum_{\kappa\sigma}\sum_{p_l}\Big(t_{\kappa p_l} c^\dagger_{\kappa\sigma_l} d_{p_l\sigma} + h.c.\Big) + \sum_{pLpR\sigma}(t_{pLpR} d^\dagger_{p_L\sigma} d_{p_R\sigma} e^{-i\theta} + h.c.) \tag{9}$$

In this final form, the mode of the resonator is coupled to the energy levels of the Left dot. Additionally, the tunneling term between the dots has acquired a phase which depends on the $\hat{p}$ operator of the mode. In the main text, the $\hat{x}$ field (describing the quantization of the electric field in the cavity) is related to $a + a^\dagger$.

We can characterize the effect of the DQD on the resonator at point A by measuring the amplitude response as a function of the driving frequency, for various bias voltages, as shown in Fig. S1. The coupling to the electronic system leads to a renormalization of the bare resonator frequency $\omega_0$ to a voltage-dependent value $\omega_0^*(V_{sd})$, which can be seen in the amplitude response of the cavity (and associated with the red dashed line in Fig. S2) [12].

**Sec. III Description in relation with an SU(4) Kondo model**

The low-energy theory of the electronic degrees of freedom at point A can be built in perturbation theory, by restricting the electron dynamics on the dot to the two allowed states $(N-1, M)$ and $(N, M-1)$. We can then introduce a pseudo-spin degree of freedom $\vec{T}$ to describe the two charge states by analogy with a double-well problem, assuming the limit of small quantum dots [13, 14]. The spin $\vec{S}$ of electrons will couple to the pseudo-spin degrees of freedom when taking into account second-order tunneling processes.

**Sec. III. A Kondo SU(4) Hamiltonian**

As a result of second-order tunneling processes, the electron leads are coupled through the Hamiltonian $H_1 = H_{kin} + H_{kondo} + H_{assist} + H_{orbital}$ [13]. Here, $H_{kin}$ represents the kinetic energy in the two leads. Second-order tunneling processes are classified through purely Kondo terms involving spin flips, orbital contributions changing the lead index from say $l = L$ (Left) to $l = R$ (Right) and flipping the charge state on the DQD, and assisted tunneling processes entangling the charge and spin [13]:

$$H_{kondo} + H_{assist} = \frac{J}{2}\vec{S}\psi^\dagger\vec{\sigma}\psi + Q_z T^z \vec{S}\psi^\dagger\tau^z\vec{\sigma}\psi + Q_\perp(T^\dagger\vec{S}\psi^\dagger\tau^-\vec{\sigma}\psi + h.c.),$$

$$H_{orbital} = \frac{1}{2}(V_z T^z\psi^\dagger\tau^z\psi + V_\perp(T^\dagger\psi^\dagger\tau^-\psi + h.c.)), \qquad (10)$$

where $\psi_{\sigma l}$ is the Fourier transform of $c_{\kappa\sigma l}$ which refers to an electron operator with spin $\sigma$ in lead $l$. Here, $\vec{S}$ corresponds to the spin operator of the electron delocalized on the DQD, $\vec{\sigma}$ and $\vec{\tau}$ act on the spin space and orbital (lead) space of the electron operators. The couplings above can be written in terms of the tunneling rate to the $\Gamma_\pm$ and the charging energy $Ec$ of the DQD respectively as [13] $J = Q_z = (\Gamma_+ + +\Gamma_-)/4E_c$, $Q_\perp = V_\perp = \sqrt{\Gamma_+\Gamma_-}/E_c$.

To generalize this Hamiltonian in graphene, we assume that an electron in graphene leads, has an equal probability to hop onto the DQD when belonging to sub-lattice $A$ or sub-lattice $B$. In this sense, the c-operator corresponds to the symmetric combination of $A$ and $B$ sub-lattice electron operators in graphene. A small asymmetry between these tunneling elements (smaller than the Kondo energy scale) should not affect the results in the main text. We argue that a small geometrical asymmetry of the tunneling between the two graphene sub-lattices is unimportant, as the symmetric term would dominate in the renormalization procedure of Ref. [13].

It should also be noted that the tunneling term between the two dots also involve the phase (operator) $\theta$ after the unitary transformation done above. Therefore, the coupling between the two dots gives rise to an orbital magnetic field $t' exp(-i\theta) T^\dagger + h.c.$. In the main text, when focusing on DC transport we redefine $t' \rightarrow t' exp(-i\theta)$. The important point is that, in contrast to a macroscopic environment which may suppress the DC conductance [14], the coupling with a single mode resonator should not affect the transmission between the two dots in the limit of zero frequency. The main process for DC conductance should not be accompanied by the emission of a (single) photon.

**Sec. III. B Fermi liquid picture**

The starting point of our analysis is that the system flows to a Kondo-Fermi-liquid fixed point [13, 15-20], quite robust to (charge) noise effects [14]. Experiments have reported the possible occurrence of an emergent $SU(4)$ symmetry at the low-energy fixed point, in similar geometries, due to the entanglement between spin and orbital degrees of freedom [21-25]. The remarkable fact is that the bare parameters of the model are replaced by a single parameter describing the low-energy fixed point, namely the Kondo energy scale $k_B T_K$. In this description the spectral function of the DQD can be modeled by an effective resonant level model [16], associated with spinless fermionic operators $d$ and $d^\dagger$ defined in the main text and describing the two macroscopic charge states on the DQD. These operators $d$ and $d^\dagger$ are defined in correspondence with the processes involving the orbital operators $T^-$ and $T^\dagger$ respectively. This spinless description is legitimate as the two spin components add up in the conductance, and the cavity only couples to the orbital degree of freedom. We stress that even though we describe the effective resonant level with spinless operators, the spin of the electrons is a highly relevant quantity: the joint effect of the orbital and spin degrees of freedom lead to a $SU(4)$ Kondo resonance at $\epsilon_0$ of the order of $k_B T_K$, above the Fermi surface.

In this regime, the system can then be treated by analogy with the case of one single dot studied in Ref. [12], the only difference being that the position $\epsilon_0$ and the width $\Gamma(V_{sd})$ of the electronic level are determined by the $SU(4)$ fixed point, leading to a the spectral function of the form :

$$\rho(\omega) = \frac{1}{\pi} \frac{\Gamma(V_{sd})}{(\hbar\omega - \epsilon_0)^2 + \Gamma(V_{sd})^2} \qquad (11)$$

where the position of the resonance $\epsilon_0$ is of the order of $k_B T_K$ [16, 20]. We take a bias dependent width $\Gamma(V_{sd})$ for the pseudo-fermion $d$ to account for dissipation effects at large voltages, see below.

**Sec. III. C Voltage dependent level width**

We estimate perturbatively the decoherence rate $\Gamma$ at large bias voltages, adopting the pseudo-fermion $d$ and $d^\dagger$ resonance description (related to the orbital degrees of freedom $T^-$ and $T^\dagger$ respectively). The decoherence channels on the Kondo effect at large bias voltages will be provided by the terms involving $T^-$ and $T^\dagger$, namely $V_\perp$ and $Q_\perp$. It is quite straightforward to compute the imaginary-part of the self energy of the $d$ fermion to second order in $V_\perp$ and $Q_\perp$.

For the $V_\perp$ channel, by analogy with the $SU(2)$ Kondo effect [26], we find

$$\Gamma_1 = 2\int d\omega' {V_\perp}^2(\omega') f_{\omega'-\mu_L}(1 - f_{\omega'-\mu_R}). \qquad (12)$$

To simplify, here we set $\hbar = 1$. In this expression, the factor 2 comes from the two spin polarizations of an electron, $f$ is the Fermi distribution and $\mu_L$ and $\mu_R$ are the chemical potentials in the two leads. Since we consider high frequency scales of the order of $\omega' \sim eV_{sd}$ or short time scales smaller than $1/(k_B T_K)$, for the pseudo-fermion $d$ evolution in time, we have replaced $exp(-i\epsilon_0 t)$ by one since $\epsilon_0 \sim T_K$.

In a similar way, for the $Q_\perp$ channel, we find

$$\Gamma_2 = 4\int d\omega' {Q_\perp}^2(\omega') f_{\omega'-\mu_L}(1 - f_{\omega'-\mu_R}). \qquad (13)$$

The factor 4 encodes all the spin possibilities allowed by the $Q_\perp$ channel. The total decoherence rate is $\Gamma = \Gamma_1 + \Gamma_2$.

To find Eq. (4) at large bias voltages in the main text, we proceed as follows. First, based on the renormalization group equations of the $SU(4)$ Kondo model [15], we can approximate $V_\perp(\omega) \sim Q_\perp(\omega) = J(\omega)$. This leads to the following equation for $\omega' \approx eV_{sd} \gg k_B T_K$:

$$\frac{1}{4J(\omega')} \approx \log\left(\frac{eV_{sd}}{k_B T_K}\right), \qquad (14)$$

where we have used the definition of the $SU(4)$ Kondo energy scale $k_B T_K = D\exp(-1/4J(D))$ where $D$ is an ultraviolet cutoff, essentially $E_C$ in the present case. Extending this form of solutions in all the frequency window set by
e$V_{sd}$ then leads to

$$\Gamma = k_B T_K \text{ for } eV_{sd} \ll k_B T_K, \qquad (15)$$

$$\Gamma \sim eV_{sd}/\ln^2(eV_{sd}/k_B T_K) \text{ for } eV_{sd} \gg k_B T_K \qquad (16)$$

by analogy with the $SU(2)$ Kondo effect [26].

At large bias voltages, the current produces dissipation and decoherence effects on the Kondo resonance. A polynomial interpolation is performed between small and large biases, and the resulting behavior is shown in Fig. S4. As can be seen in this figure, $(\Gamma(V_{sd}) - eV_{sd})$ (given by the space between the magenta line and the dashed black line) initially decreases until e$|V_{sd}| \sim k_B T_K$ which is responsible for the low-energy features in the amplitude. For e$|V_{sd}| \geq k_B T_K$, we still observe a robust $\pi$ phase in the reflected signal. At e$|V_{sd}|/k_B T_K \sim 10$, the quantity $\Gamma(V_{sd})$ becomes smaller than $V_{sd}$, leading to a phase shift from $\pi$ to 0. Above e$|V_{sd}| \sim E_C$ (vertical dotted blue line) the model ceases to be valid as we should take into account other energy levels. As $T_K$ can be estimated from DC transport measurements (see Fig. S3), we only use one fitting parameter to obtain Fig. S4: the proportionality coefficient on the right hand side of Eq.

(16); we also check that changing the dissipation strength $\alpha$ does not affect much the results (in the weak-dissipation regime).

This particular evolution of the resonance width $\Gamma$ determines the electron-induced photon lifetime, $\text{Im}\,\Pi^R(\omega_0^*(V_{sd}))$. Using the Fermi liquid picture then gives [12]:

$$\text{Im}\,\Pi^R(\omega_0^*) = \lambda^2 f_\Gamma(\omega_0^*) \sum_{\alpha,a=\pm} \alpha \arctan\left(\frac{\mu_a-\epsilon_0+\alpha\hbar\omega_0^*}{\Gamma}\right) +$$

$$\lambda^2 f_\Gamma(\omega_0^*) \sum_{\alpha,a=\pm} \frac{\Gamma}{\omega_0^*} \ln\left(\frac{(\mu_a-\epsilon_0+\alpha\hbar\omega_0^*)^2+\Gamma^2}{(\mu_a-\epsilon_0)^2+\Gamma^2}\right), \qquad (17)$$

where $f_\Gamma(\omega_0^*) = \Gamma/(4\pi^2\Gamma^2 + \pi^2\hbar^2\omega_0^*)$. In addition, $\mu_a = aeV_{sd}/2$ are the chemical potentials associated with each lead, where formally $a = +1$ for left (L) lead and $a = -1$ for right (R) lead. Formally, this form of $\text{Im}\,\Pi^R(\omega_0^*)$ is obtained by considering the limit $|t'|\lambda/\omega_0^* \ll \lambda$.

The phase and amplitude of transmitted photons are then directly obtained from $\text{Im}\,\Pi^R(\omega_0^*)$, with Eqs. (1) and (2) of the main text. Corresponding results are shown in Fig. 2 of the main text.

**Sec. IV. Coulomb blockade regime and input power dependence**

We now study the power ($P$) dependence of the resonator response as a function of the bias voltage in the Coulomb blockade regime at point D. Fixing the driving frequency at resonance $\omega = \omega_0^*(V_{sd})$, we show the amplitude response (figures S5(a) and (c)) and the phase response (figures S5(b) and (d)) for two different values of the power. Figures S5(a) and (b) show the phase and amplitude response at weak power. A phase shift is confirmed from bias voltage $V_{sd} = 0$ to $V_{sd} = \pm 0.5$ mV. The amplitude sharply goes down for about $-15$ dB from $V_{sd} = 0$ to $V_{sd} = \pm 0.1$ mV, and two side peaks appear around $V_{sd} = \pm 0.15$ mV. Then, it continues to go down and two dips appear around $V_{sd} = \pm 0.4$ mV, which are again consistent with the charging energy $E_C$ of the quantum dots measured by the Coulomb diamond. At small $V_{sd}$, this confirms that the Coulomb blockade phenomenon pins the charge fluctuations on the DQD and as a result the photon field is weakly affected by the matter. When the bias voltage increases and compensates the charging energy, there is a revival of charge fluctuations on the DQD. It is important to note that the cavity is only coupled to the orbital degree of freedom, leading to a phase of zero when $V_{sd} \to 0$. As a result, it is not sensitive to an eventual spin Kondo effect where the spin and orbital degrees of freedom are decoupled (resulting from the system composed of the left dot and the left lead alone). A more refined theoretical analysis would be necessary to describe quantitatively the crossover with voltage and to include the charging energy effects.

We see that the phase response is drastically modified as the power changes. At high power the amplitude and phase response are close to the ones obtained at point A. This observation can be explained by the fact that the photonic interaction acts as an artificial chemical potential at high drive powers, bringing the system from the Coulomb blockade regime to the resonant regime. More precisely, in the large drive regime, we can decompose the photon field as the sum of its mean value plus quantum fluctuations. This leads to a rewriting of the photon annihilation operator as $a = \langle a \rangle +$ $d$. The first part $\langle a\rangle$ is determined by the strength of the drive field and the damping rate, and we suppose that steady state of the cavity is a coherent state. The constant semi-classical part of the drive $\langle a+a^\dagger\rangle$ acts as a chemical potential on the Left dot. Following Ref. [2] the chemical potentials of the two dots $\mu_L$ and $\mu_R$ read as a function of the number of electrons on the dots ($N_L$ and $N_R$),

$$\mu_L(N_L,N_R)=\left(N_L-\frac{1}{2}-\lambda\langle a+a^\dagger\rangle\right)E_{C_L}+N_R E_{C_m}-\frac{1}{|e|}(C_{gL}V_{LP}E_{C_L}+C_{gR}V_{RP}E_{C_m}), \quad (18)$$

$$\mu_R(N_L,N_R)=\left(N_R-\frac{1}{2}\right)E_{C_R}+N_L E_{C_m}-\frac{1}{|e|}(C_{gL}V_{LP}E_{C_m}+C_{gR}V_{RP}E_{C_R}), \quad (19)$$

where $E_{C_j}$ is the charging energy of the individual dot $j$, $E_{C_m}$ is the electrostatic coupling energy, $C_j$ ($C_{gj}$) is the capacitance coupling the dot $j$ to the neighboring lead (gate), and $C_m$ is the capacitance which couples the dots together. The Left and Right gate voltages are denoted by $V_{LP}$ and $V_{RP}$. The effect of the driving can be absorbed into the definition of new gates voltages $V'_{LP}$ and $V'_{RP}$,

$$V'_{LP}=V_{LP}-\frac{\lambda\langle a+a^\dagger\rangle|e|}{C_{gL}(E_{C_L}-\frac{E_{C_m}^2}{E_{C_R}})}, \quad (20)$$

$$V'_{RP}=V_{RP}-\frac{\lambda\langle a+a^\dagger\rangle|e|}{C_{gR}(E_{C_m}-\frac{E_{C_L}E_{C_R}}{E_{C_m}})}. \quad (21)$$

Driving the system at high power allows to move the state of the system on the honeycomb phase diagram along a line characterized by the equations (20) and (21). This corresponds to

$$\frac{\delta V_{RP}}{\delta V_{LP}}=\frac{C_{gL}}{C_{gR}}\frac{E_{C_m}}{E_{C_R}}. \quad (22)$$

Such a line has a slope which is the opposite of the line of degeneracy between the states $(N,M-1)$ and $(N,M)$ (the line where lies the point B).

We can see from Fig. S6 that driving the cavity may bring the system at resonance. We can estimate the input power that allows to make this transition from the Coulomb Blockade regime to the resonant regime. In the steady state, the input-output theory allows to relate the drive power to the number of photons in the cavity and ultimately to $\langle a+a^\dagger\rangle$. We have $P=(\hbar\omega\eta/4)\langle a^\dagger a\rangle$, where $\eta$ is the decay rate of the photons in the leads (assumed to be independent over the range of frequencies relevant to the cavity in the Markov approximation). If we suppose that the steady state of the cavity is a coherent state, we have

$$\langle a+a^\dagger\rangle \cong \sqrt{\frac{P}{\hbar\omega\eta}}. \quad (23)$$

We conclude that, at a mean field level, driving the system indeed allows to move the state of the system on the honeycomb phase diagram, as the mean field contribution can shift the left chemical potential. In the steady state, the input-output theory allows to relate the drive power to the number of photons in the cavity and ultimately to $\langle a + a^\dagger \rangle$ if we suppose that the steady state of the cavity is a coherent state. Following Ref. [27], we have $P = \hbar\omega \frac{\eta}{4} \langle a^\dagger a \rangle$. For a coherent state in the cavity this allows to write $\langle a + a^\dagger \rangle \cong \sqrt{P/\hbar\omega\eta}$. We can then estimate the critical power $P_c$ needed to compensate $E_C$. We have $E_C = 0.4$ meV, $\omega = 7$ GHz, and we take $\lambda \cong 0.15$ GHz and $\eta \cong 0.1$ GHz. This finally gives $P_c \cong 10^{-10}$ V $= -70$ dBm, which is consistent with the experimental observations. We note that the high-power responses at small biases differ slightly from those at point A. This may be linked to additional non-equilibrium effects arising at high power.

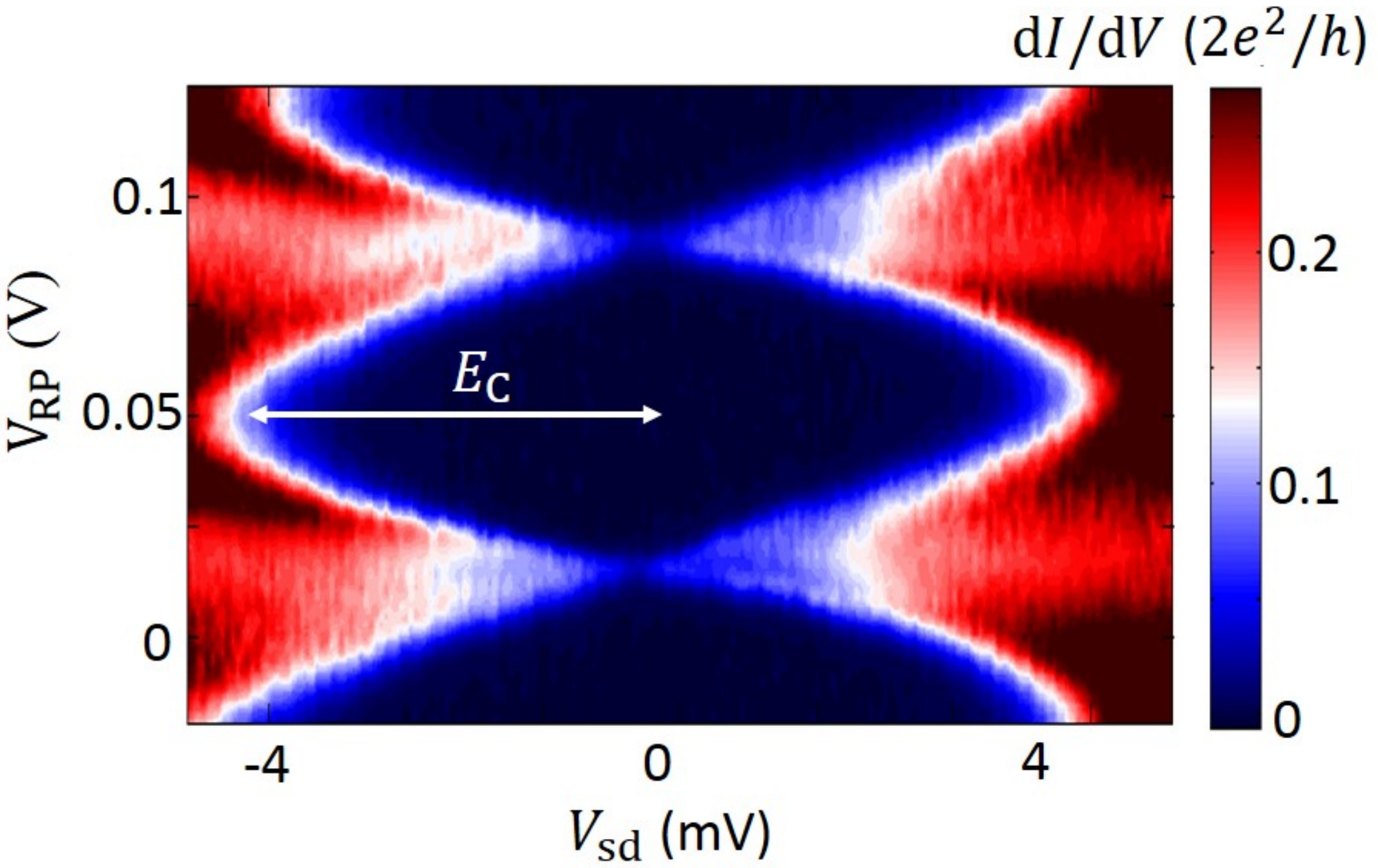

**Fig. S1. Coulomb diamond of the quantum dot.** Coulomb diamond measured by transport experiment and revealing the charging energy $E_C$. We use a single charging energy in agreement with previous measurements [7, 8]. The charging energies of the two dots are measured to be almost identical, and we find that the mutual capacitance is of the order of the single dot capacitances. All microwave and DC transport measurements support this argument. Gate lever arms are ~10%.

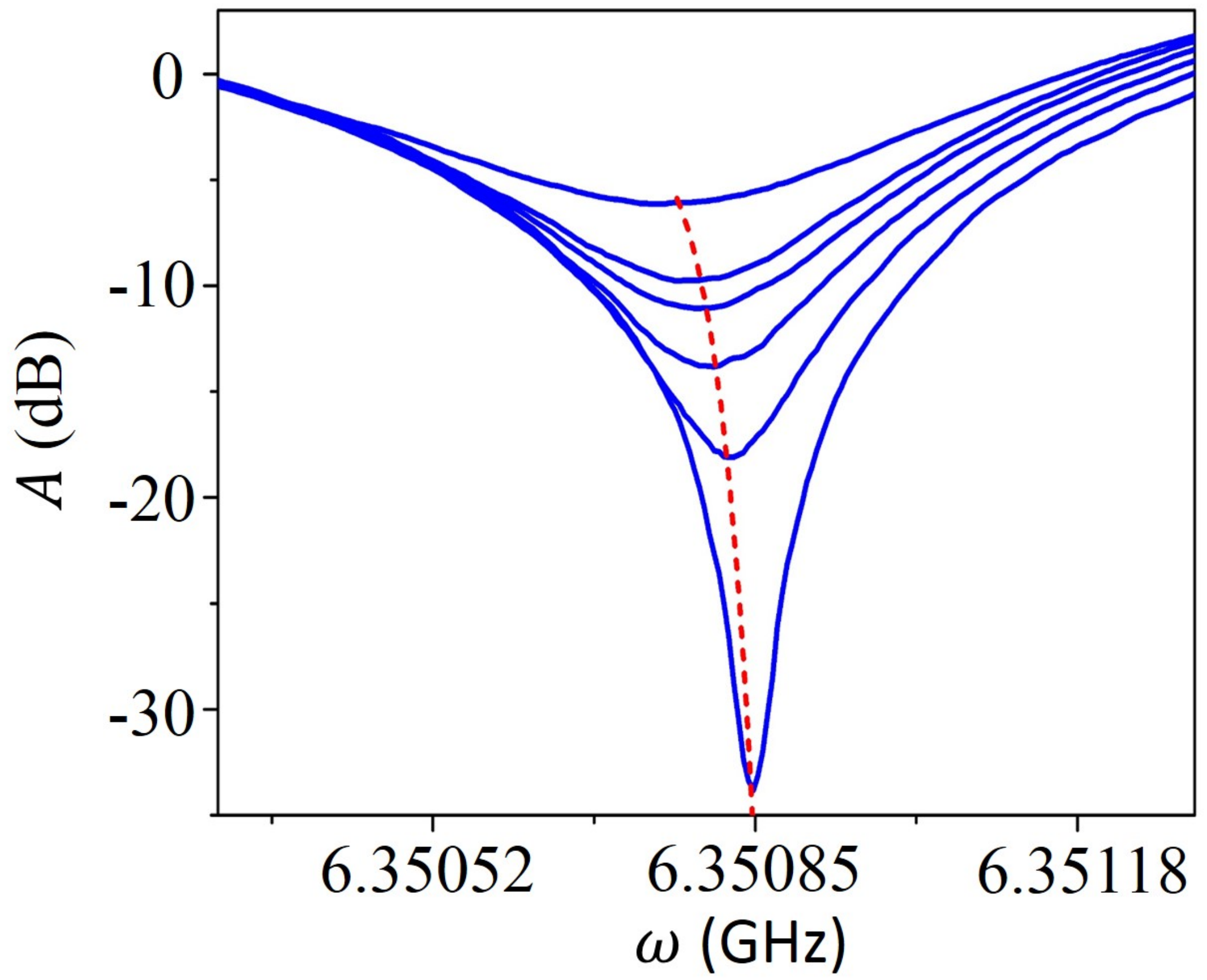

**Fig. S2. Amplitude response as a function of the driving frequency**. Amplitude response $A$ as a function of the driving frequency $\omega$, for various bias voltages (0, 0.5, 0.6, 0.7, 0.8, 1 mV) at point A. The red dashed line shows the fitted resonance frequency shifted by the electron transport, which allows us to determine the light matter coupling $\lambda/(\hbar\omega_0) \cong 10^{-2}$ (see main Text). This small value justifies a perturbative calculation in $\lambda$.

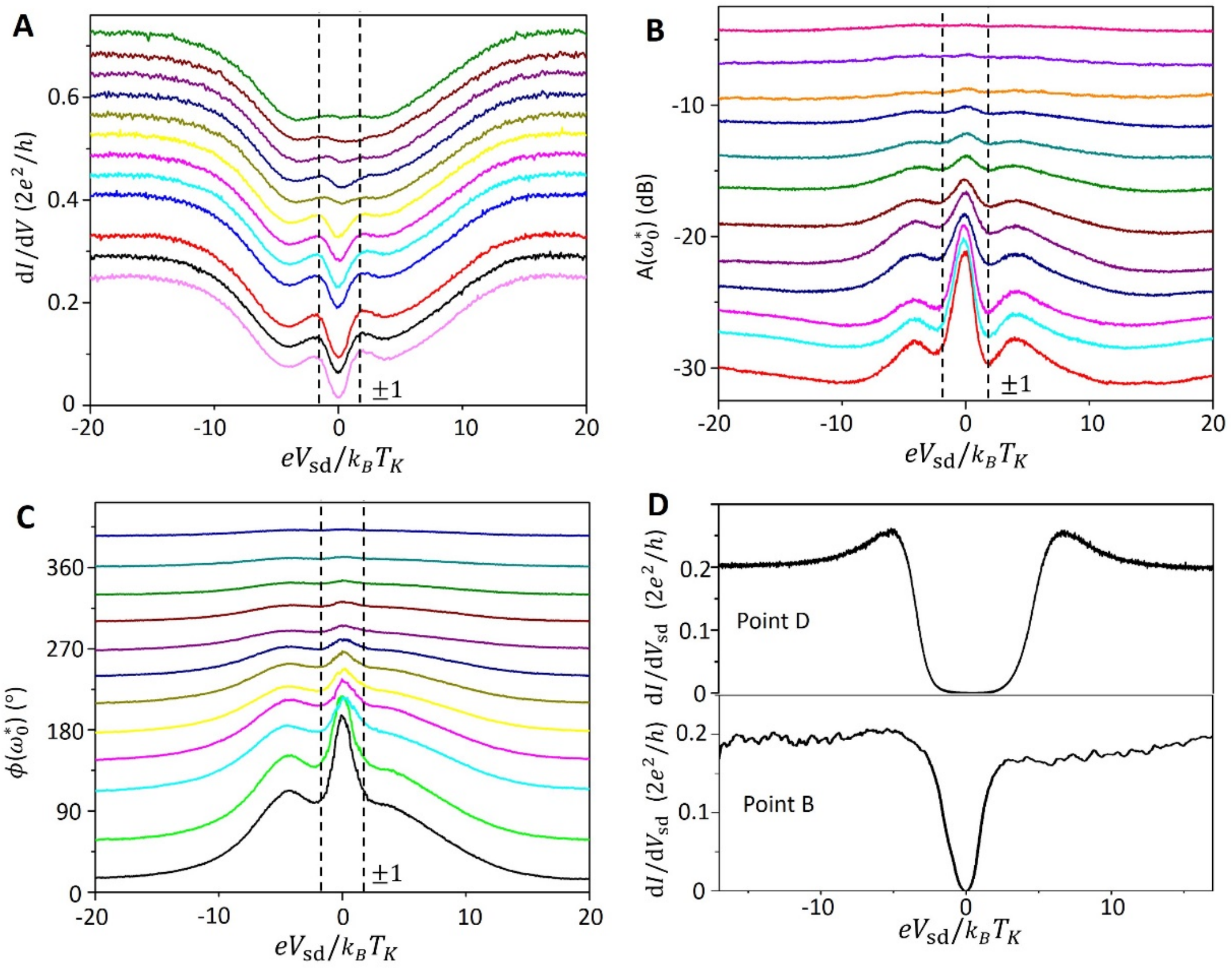

**Fig. S3. Temperature dependence.** Transport differential conductance (A), microwave amplitude (B) and phase (C) at point A in Fig. 1 D, showing two dips (peaks) at finite voltage $eV_{sd} \sim \pm k_BT_K$ which smoothly disappear above a temperature of about 550 mK. Curves are displaced from the base reference by a constant offset, the temperatures are 50, 100, 200, 300, 400, 500, 550, 600, 650, 700, 750, and 800 mK from bottom to top. Interestingly, the phase converges to π for $T<T_K$=550 mK for $V_{sd} \rightarrow 0$ and its evolution reveals the Kondo temperature. In addition, for intermediate temperatures, the phase smoothly drops to zero for $eV_{sd} > k_BT_K$. For an SU(4) Kondo model in this configuration, the low-temperature conductance at $V_{sd} \rightarrow 0$ is expected to follow the small transmission probability between the two dots, and is therefore much smaller than $2e^2/h$. Fitting with the theory and using $T_K$=550 mK together with Eq. (S11), then we find $t' \sim 0.3\ k_BT_K$. Small peaks occur at voltages of the order of $k_BT_K$ in agreement with theory (*23*). (D) Transport differential conductance at point D (upper) and point B (lower, conductance at point C is very similar with point B shown here) at $T$=30 mK.

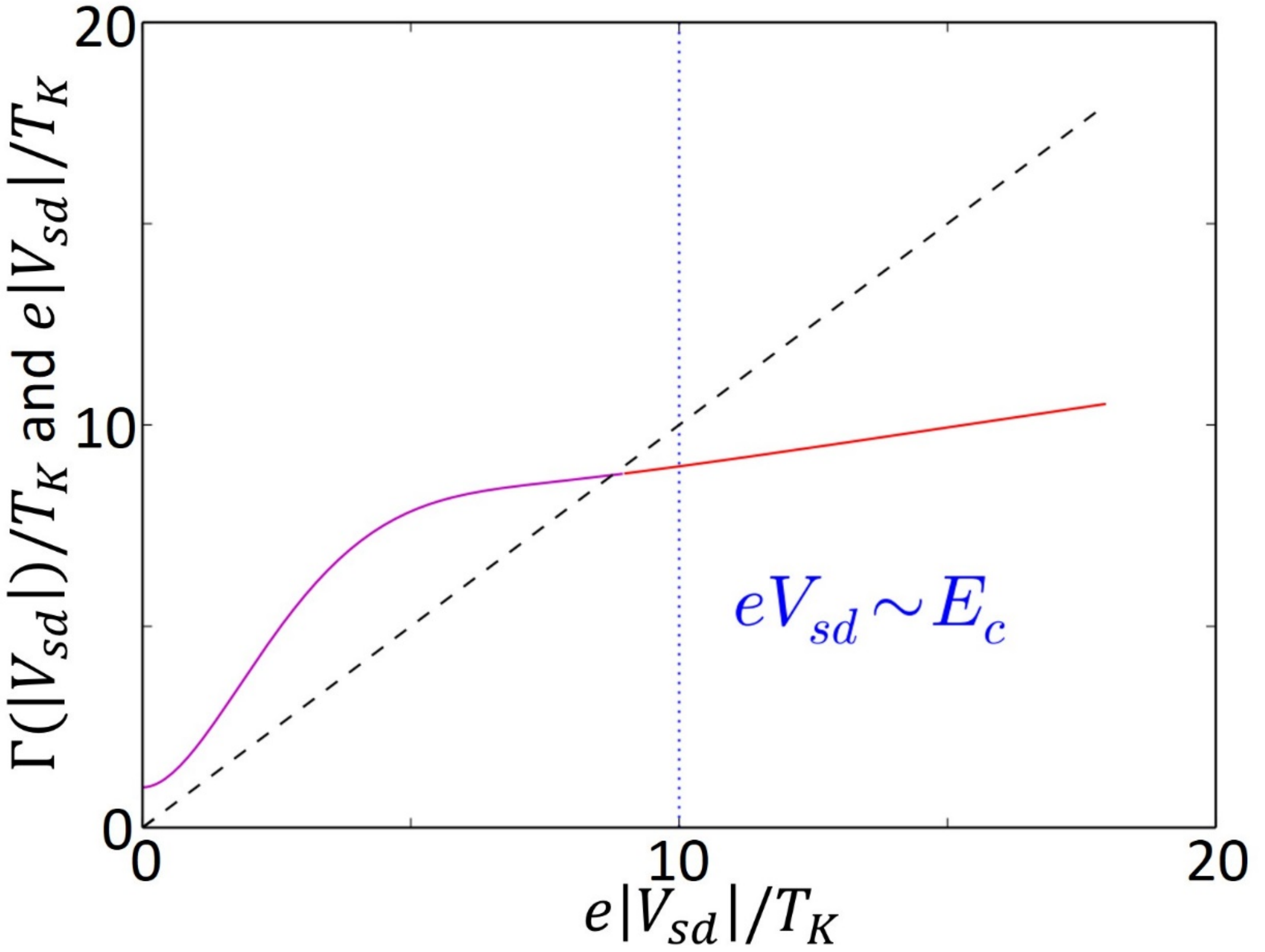

**Fig. S4. Evolution of $\Gamma$ with respect to bias voltages.** Evolution of $\Gamma/T_K$ with respect to $eV_{sd}/T_K$ (here $k_B$ =1). At low bias voltages, we have $\Gamma/T_K{\sim}1$, while $\Gamma{\sim}V_{sd}[\ln(V_{sd}/T_K)]^2$ for large biases $eV_{sd}/T_K \gg 1$ (full red line). This expression is valid asymptotically and for $\Gamma < V_{sd}$ [26]. We use a polynomial (here degree 6) interpolation between these two limits (full magenta line). The vertical dotted blue line shows the value of the charging energy $E_C$.

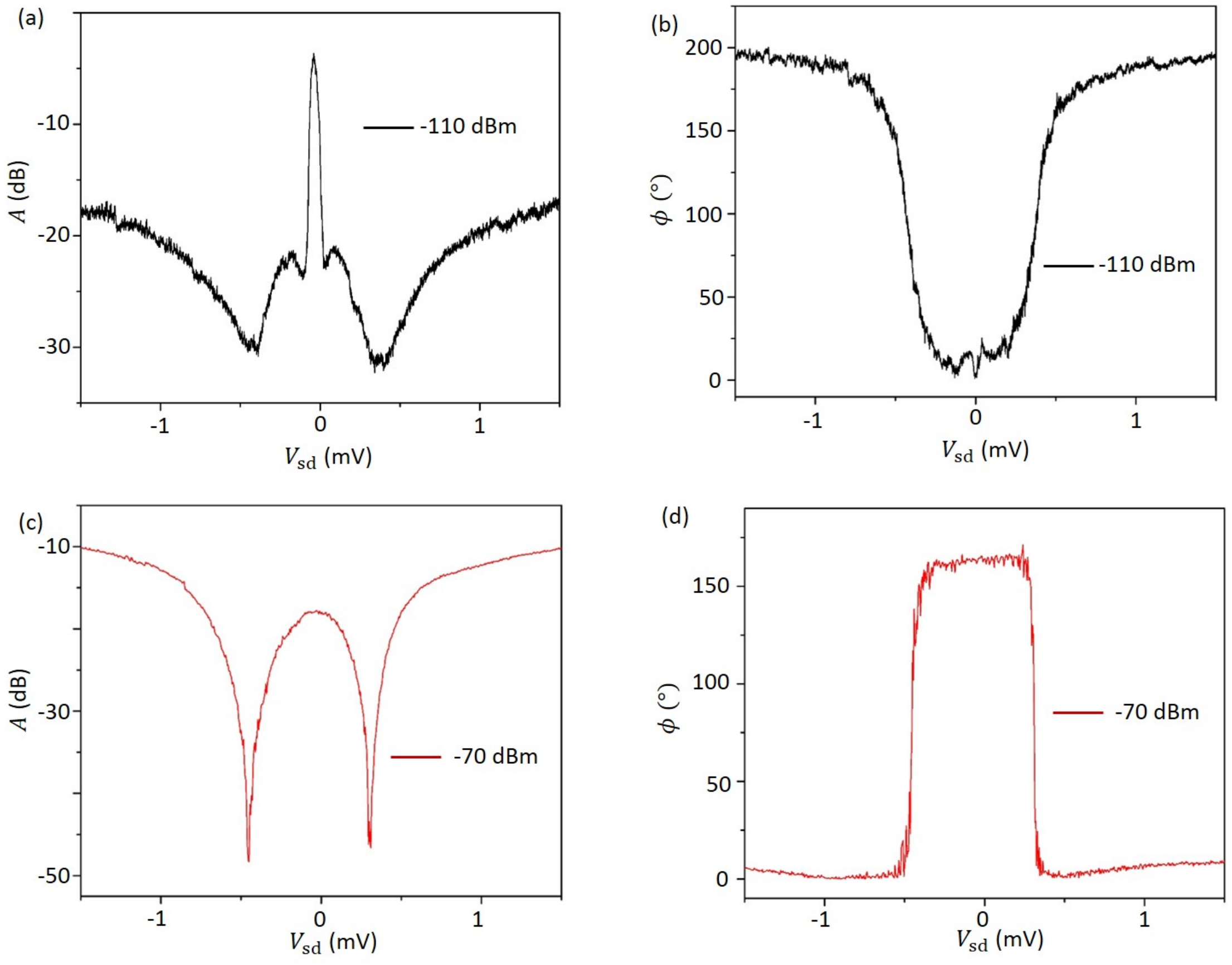

**Fig. S5. Microwave response as a function of bias voltages when increasing the driving power**. Amplitude (a) and phase (b) response as functions of the bias voltage for the point D. Same results when increasing the microwave power (c) and (d).

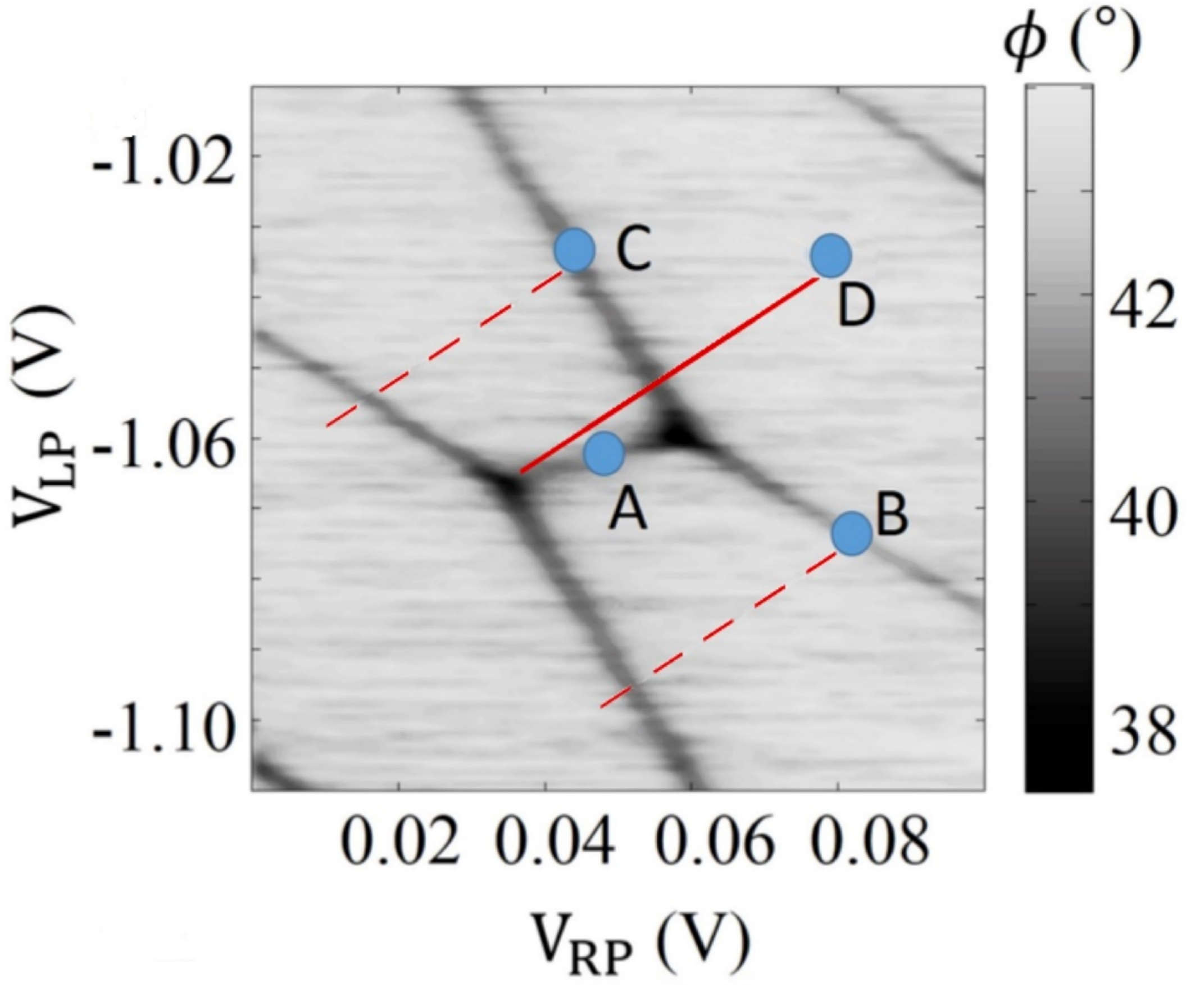

**Fig. S6. Effect of the driving on the DQD states.** The full red line shows the shift in the phase diagram induced by the drive at point D.

**SI Reference**

[1] D. Wei, H.-O. Li, G. Cao, G. Luo, Z.-X. Zheng, T. Tu, M. Xiao, G.-C. Guo, H.-W. Jiang, and G.-P. Guo, “Tuning inter-dot coupling of an etched graphene double quantum dot by adjacent metal gates,” Sci. Rep. 3, 3175 (2013).

[2] W. G. van der Wiel, S. De Franceschi, J. M. Elzerman, T. Fujisawa, S. Tarucha, and L. P. Kouwenhoven, “Electron transport through double quantum dots,” Rev. Mod. Phys. 75, 1 (2002).

[3] R. Ziegler, C. Bruder, and H. Schoeller, “Transport through double quantum dots,” Phys. Rev. B 62, 1961 (2000).

[4] A. H. Castro Neto, F. Guinea, N. M. R. Peres, K. S. Novoselov, and A. K. Geim, “The electronic properties of graphene,” Rev. Mod. Phys. 81, 109–162 (2009).

[5] R. Egger and O. Gogolin, “Correlated transport and non-fermi-liquid behaviour in single-wall carbon nanotubes,” Eur. Phys. J. B 3 (3), 281–300 (1998).

[6] K. Le Hur, S. Vishveshwara, and C. Bena, “Double-gap superconducting proximity effect in armchair carbon nanotubes,” Phys. Rev. B 77, 041406 (R) (2008).

[7] G.-W. Deng, D. Wei, J.R. Johansson, M.-L. Zhang, S.-X. Li, H.-O. Li, G. Cao, M. Xiao, T. Tu, G.-C. Guo, H.-W. Jiang, F. Nori, and G.-P. Guo, “Charge number dependence of the dephasing rates of a graphene double quantum dot in a circuit QED architecture,” Phys. Rev. Lett. 115, 126804 (2015).

[8] G.-W. Deng, D. Wei, S.-X. Li, J. R. Johansson, W.-C. Kong, H.-O. Li, G. Cao, M. Xiao, G.-C. Guo, F. Nori, H.-W. Jiang, and G.-P. Guo, “Coupling two distant double quantum dots with a microwave resonator,” Nano Letters 15, 6620–6625 (2015).

[9] L. Childress, A. Sorensen, and M. Lukin, “Mesoscopic cavity quantum electrodynamics with quantum dots,” Phys. Rev. A 69, 042302 (2004).

[10] K. D. Petersson, L. W. McFaul, M. D. Schroer, M. Jung, J. M. Taylor, A. A. Houck, and J. R. Petta, “Circuit quantum electrodynamics with a spin qubit,” Nature 490, 380–383 (2012).

[11] A. Cottet, T. Kontos, and B. Doucot, “Electron-photon coupling in mesoscopic quantum electrodynamics,” Phys. Rev. B 91, 205417 (2015).

[12] M. Schiró and K. Le Hur, “Tunable hybrid quantum electrodynamics from nonlinear electron transport,” Phys. Rev. B 89, 195127 (2014).

[13] L. Borda, G. Zarand, W. Hofstetter, B.I. Halperin, and J. von Delft, “SU(4) fermi liquid state and spin filtering in a double quantum dot system,” Phys. Rev. Lett. 90, 026602 (2003).

[14] Mei-Rong Li and Karyn Le Hur, “Double-dot charge qubit and transport via dissipative cotunneling,” Phys. Rev. Lett. 93, 176802 (2004).

[15] K. Le Hur, P. Simon, and L. Borda, “Maximized orbital and spin Kondo effects in

a single-electron transistor," Phys. Rev. B 69, 045326 (2004).
[16] K. Le Hur, P. Simon, and D. Loss, "Transport through a quantum dot with SU(4) Kondo entanglement," Phys. Rev. B 75, 035332 (2007).
[17] R. Lopez, D. Sanchez, M. Lee, M.-S. Choi, P. Simon, and K. Le Hur, "Probing spin and orbital Kondo effects with a mesoscopic interferometer," Phys. Rev. B 71, 115312 (2005).
[18] P. Nozi è res, "A 'fermi-liquid' description of the kondo problem at low temperatures," J. Low Temp. Phys. 17, 31–42 (1974).
[19] I. Affleck and A. Ludwig, "Exact conformal-field-theory results on the multichannel Kondo effect: Single-fermion green's function, self-energy, and resistivity," Phys. Rev. B 48, 7297 (1993).
[20] C. Mora, P. Vitushinsky, X. Leyronas, A. A. Clerk, and K. Le Hur, "Theory of non-equilibrium transport in the SU(N) Kondo regime," Phys. Rev. B 80, 155322 (2009).
[21] A. J. Keller, S. Amasha, I. Weymann, C. P. Moca, I. G. Rau, J. A. Katine, Hadas Shtrikman, G. Zarand, and D. Goldhaber-Gordon, "Observation of the SU(4) Kondo state in a double quantum dot," Nature Phys. 10, 145–150 (2014).
[22] A. Makarovski, J. Liu, and G. Finkelstein, "Evolution of SU(4) transport regimes in carbon nanotube quantum dots," Phys. Rev. Lett. 99, 066801 (2007).
[23] S. Sasaki, S. Amaha, N. Asakawa, M. Eto, and S. Tarucha, "Enhanced Kondo effect via tuned orbital degeneracy in a spin-1/2 artificial atom," Phys. Rev. Lett. 93, 017205 (2004).
[24] T. Delattre, C. Feuillet-Palma, L.G. Herrmann, P. Morfin, J.-M. Berroir, G. Fève, B. Placais, D.C. Glattli, M.-S. Choi, C. Mora, and T. Kontos, "Noisy Kondo impurities," Nature Physics 5, 208 (2009).
[25] P. Jarillo-Herrero, J. Kong, H. S.J. van der Zant, C. Dekker, L. P. Kouwenhoven, and S. De Franceschi, "Orbital Kondo effect in carbon nanotubes," Nature 434, 484 (2005).
[26] A. Rosch, J. Kroha, and P. Wölfle, "The Kondo effect in quantum dots at high voltage: Universality and scaling," Phys. Rev. Lett. 87, 156802 (2001).
[27] A.A. Clerk, M.H. Devoret, S.M. Girvin, F. Marquardt, and R.J. Schoelkopf, "Introduction to quantum noise, measurement and amplification," Rev. Mod. Phys. 82, 1155 (2010).